\documentclass[referee]{aa} 
\usepackage{graphicx}
\usepackage{natbib}
\begin{document}
   \title{Photodesorption of ices I: CO, N$_2$ and CO$_2$}

   \author{Karin I. \"Oberg\inst{1}
          \and
          Ewine F. van Dishoeck\inst{2,3}
		  \and
		  Harold Linnartz\inst{1}
          }

   \institute{Raymond and Beverly Sackler Laboratory for Astrophysics, Leiden Observatory, Leiden University, P.O. Box 9513, NL 2300 RA Leiden, The Netherlands.\\
              \email{oberg@strw.leidenuniv.nl}
         \and
             Leiden Observatory, Leiden University, P.O. Box 9513, NL 2300 RA Leiden, The Netherlands
         \and
             Max-Planck-Institut f\"ur extraterrestrische Physik (MPE), Giessenbachstraat 1, 85748 Garching, Germany\\
             }

   \date{Received; accepted}

 
  \abstract
   {A longstanding problem in astrochemistry is how molecules can be maintained in the gas phase in dense inter- and circumstellar regions at temperatures well below their thermal desorption values. Photodesorption is a non-thermal desorption mechanism, which may explain the small amounts of observed cold gas in cloud cores and disk mid-planes. }
   {This study aims to determine the UV photodesorption yields and to constrain the photodesorption mechanisms of three astrochemically relevant ices: CO, N$_2$ and CO$_2$. In addition, the possibility of co-desorption in mixed and layered CO:N$_2$ ices is explored.}
   {The UV photodesorption of ices is studied experimentally under ultra high vacuum conditions and at astrochemically relevant temperatures (15 -- 60~K) using a hydrogen discharge lamp (7--10.5~eV). The ice desorption is monitored by reflection absorption infrared spectroscopy of the ice and simultaneous mass spectrometry of the desorbed molecules.}
   {Both the UV photodesorption yield per incident photon and the photodesorption mechanism are highly molecule specific. The CO photodesorbs without dissociation from the surface layer of the ice, and N$_2$, which lacks a dipole allowed electronic transition in the wavelength range of the lamp, has a photodesorption yield that is more than an order of magnitude lower. This yield increases significantly due to co-desorption when N$_2$ is mixed in with, or layered on top of, CO ice. CO$_2$ photodesorbs through dissociation and subsequent recombination from the top 10 layers of the ice. At low temperatures (15 -- 18~K), the derived photodesorption yields are $2.7(\pm1.3)\times10^{-3}$ and $<2\times10^{-4}$ molecules photon$^{-1}$ for pure CO and N$_2$, respectively. The CO$_2$ photodesorption yield is $1.2(\pm0.7)\times10^{-3}\times (1-e^{-(x/2.9(\pm1.1) \rm)})+1.1(\pm0.7)\times10^{-3}\times(1- e^{-(x/4.6(\pm2.2)} \rm))$ molecules photon$^{-1}$, where $x$ is the ice thickness in monolayers and the two parts of the expression represent a CO$_2$ and a CO photodesorption pathway, respectively. At higher temperatures, the CO ice photodesorption yield decreases, while that of CO$_2$ increases.}
   {}


   \maketitle
%

\section{Introduction}

In dark clouds molecules and atoms collide with and stick to cold submicron-sized dust particles, resulting in icy mantles \citep{Leger85, Boogert04}. The ices are subsequently processed by atom or light interactions to form more complex species \citep{Tielens82, Watanabe03,Ioppolo08}. Observations show that H$_2$O, CO and CO$_2$ are the main ice constituents, with abundances up to $10^{-4}$ with respect to the total hydrogen density. These molecules are key constituents in the formation of more complex species \citep{Tielens97}, and their partitioning between the grain and gas phase therefore strongly affects the chemical evolution in star- and planet-forming regions \citep{Vandishoeck06b}. 

Whether formed on the grains or frozen out from the gas phase, chemical models of cloud cores show that all molecules except for H$_2$ are removed from the gas phase within $\sim 10^9 / n_{\rm H}$ years, where $n_{\rm H}$ is the total hydrogen number density \citep{Willacy98}. For a typical cloud core density of $10^4$ cm$^{-3}$, this time scale is much shorter than the estimated age of such regions and thus molecules like CO and CO$_2$ should be completely frozen out. Yet gas-phase molecules, like CO, are detected in these clouds \citep{Bergin01, Bergin02}. Cold CO gas is also detected in the midplanes of protoplanetary disks \citep{Dartois03, Pietu07} where the densities are higher and the freeze-out time scales are even shorter, suggesting the existence of either efficient non-thermal desorption or an efficient mixing process in the disks. Similarly \citet{Sakai08} have detected cold HCO$_2^+$, tracing gas phase CO$_2$, toward the embedded low-mass protostar IRAS 04368+2557 in L 1527 also referred to as L 1527 IRS. From the high column density and the thin line profile they conclude that the observed CO$_2$ cannot originate from thermal evaporation of ices in the hot inner regions of the envelope. They instead suggest gas phase formation of CO$_2$ to explain their observations, but do not consider non-thermal desorption in the cold envelope as an alternative. HCO$_2^+$ is also detected by \citet{Turner99} toward several small translucent molecular clouds. They conclude that the observed HCO$_2^+$ can only form through gas phase chemistry for very specific C/O ratios and time spans and that the source of gas phase HCO$_2^+$ may instead be desorbed CO$_2$ ice. Both the CO and CO$_2$ observations may thus be explained by non-thermal desorption of ices, but this has not been quantified to date.

In dense clouds and in outer disks and disk midplanes, desorption must occur non-thermally since the grain temperature is low enough, around 10~K, that thermal desorption is negligible. Suggested non-thermal desorption pathways include photon and cosmic ray induced processes and desorption following the release of chemical energy \citep{Shen04, Roberts07}.The importance of these processes depend both on the intrinsic desorption yields and on the local environment, especially the UV and cosmic ray fluxes. External UV photons from the interstellar radiation field can penetrate into the outer regions of dense clouds and disks and this UV field may be enhanced by orders of magnitude in disks through irradiation by the young star. In addition to direct interaction with ices, cosmic rays and X-rays also produce a UV field inside of the clouds through interaction with H$_2$. 

UV photodesorption is therefore possible in most dense astrophysical environments, but it has been proposed as an important desorption pathway of ices mainly in protoplanetary disks and other astrophysical regions with dense clumps of material and excess UV photons \citep{Willacy00,Dominik05}. There is however a lack of experimentally determined photodesorption yields for most astrophysically relevant molecules. This has prevented progress in the field and in most models UV photodesorption is simply neglected. Recently we showed that CO photodesorption is an efficient process with a yield of $3(\pm1)\times10^{-3}$ photon$^{-1}$ \citep{Oberg07b}. This is of the same order as H$_2$O photodesorption, investigated by \citet{Westley95a,Westley95b}, though the dependence of the H$_2$O yield on different parameters remains unclear. The photodesorption of H$_2$O and benzene in a H$_2$O dominated ice has also been investigated by \citet{Thrower08} who only find substrate and matrix mediated desorption processes.

In this study we determine the photodesorption yield of CO$_2$ and its dependence on ice thickness, temperature, morphology, UV flux and integrated UV flux or fluence as well as UV irradiation time. In addition, we extend the previously reported investigation of CO and N$_2$ photodesorption to include different temperatures and ice morphologies. From the deduced yield dependencies we constrain the different desorption mechanisms and discuss the astrophysical implications.  


\section{Experiments and their analysis}

\subsection{Experimental details}

The experimental set-up (CRYOPAD) is described in detail by \citet{Fuchs06} and \citet{Oberg07b}. The set-up allows simultaneous detection of molecules in the gas phase by quadrupole mass
spectrometry (QMS) and in the ice by reflection absorption infrared
spectroscopy (RAIRS), with an angle of incidence of 84$^{\circ}$, using a Fourier transform infrared (FTIR) spectrometer. The FTIR covers 1200 -- 4000~cm$^{-1}$ with a spectral resolution of 0.5--1~cm$^{-1}$.  

In the experiments presented here, thin ices of 2.1--16.5 monolayers (ML) are grown with monolayer precision under ultra-high vacuum conditions ($P\sim10^{-10}$ mbar) at 15 -- 60~K on a gold substrate that is mounted on a He cryostat. All experiments are conducted with the isotopologues $^{13}$CO and $^{13}$C$^{18}$O  (Cambridge Isotope Laboratories 99\% and 97\% purity, respectively), $^{15}$N$_2$ (Campro Scientific 98\% purity), and $^{13}$CO$_2$ (Indugas 99\% purity) and $^{13}$C$^{18}$O$_2$ (ICON Isotopes 96\% purity) to avoid contributions from atmospheric contaminations as well as to be able to separate CO and N$_2$ mass spectrometrically with the QMS. Test experiments show that the isotopologue choices do not affect the experimental outcomes for any of the ices.

Within the uncertainties of the experiment, we also find that there is no difference in the photodesorption yield of 6.5~ML CO$_2$ ice deposited on top of another 7~ML CO$_2$ ice (of a different isotopologue), or 7~ML CO$_2$ on top of 10~ML of H$_2$O ice, compared with 6--7~ML CO$_2$ ice deposited directly onto the gold substrate. Since the character of the substrate seems to have no influence on the photodesorption process, all other experiments are carried out with CO$_2$ ices directly on top of the gold substrate.

The ice films are irradiated at normal or 45$^{\circ}$ incidence with UV light from a broadband hydrogen microwave discharge lamp, which peaks around Ly $\alpha$ at 121~nm and covers 115--170~nm or 7--10.5~eV \citep{Munozcaro03}.  All photodesorption experiments are performed in the same experimental chamber and the different UV angles of incidence are obtained by rotating the gold substrate. The lamp emission resembles the spectral distribution of the UV interstellar radiation field that impinges externally on all
clouds. It is also consistent with the UV radiation produced locally inside
clouds by the decay of electronic states of H$_2$, following excitation by energetic
electrons resulting from cosmic-ray induced ionization of hydrogen, see e.g.
\cite{Sternberg87}. 

The 45$^{\circ}$ and the normal incidence irradiation settings produce the same experimental results, except for a reduced photon flux on the ice in the 45$^{\circ}$ setting due to geometry. The lamp UV flux is varied between 1.1 and 8.3 $\times 10^{13}$ photons cm$^{-2}$ s$^{-1}$ in the different experiments. The UV flux is monitored during all experiments using the photoelectric effect in a thin gold wire in front of the lamp. Before the start of the experimental run, this gold wire current was calibrated to an absolute flux in a separate set-up by simultaneously measuring the flux with a NIST calibrated silicon diode and the current induced in the gold wire. During the photodesorption experiments the flux onto the ice surface is also estimated by measuring the CO$_2$  photodissociation cross-section several times during the experimental run, at both normal and 45$^{\circ}$ incidence, and comparing our derived cross sections with the calibrated values in \citet{Cottin03}. This was deemed necessary since the calibration measurements were carried out with normal incidence in the separate set-up, while most experiments used a 45$^{\circ}$ incidence angle. To prevent photodesorption during these measurements, the CO$_2$ ice is covered with an inert ice layer. We find that in the normal incidence setting the resulting flux using this actiometry method deviates by a factor 0.9-1.4 from the photodiode-calibrated gold-wire results.

Tables \ref{con2exps} and \ref{co2exps} summarize the experiments in this study. In the CO experiments, the temperature is varied between 15 and 27~K, which is close to its thermal desorption temperature \citep{oberg05}. This complements the previous CO photodesorption experiments, which investigated the dependences of photodesorption on lamp flux and ice thickness at 15~K \citep{Oberg07b}. In three additional experiments the photodesorption yield (or upper limits) of N$_2$ is determined, as well as the changes in CO and N$_2$ ice photodesorption in a CO:N$_2$ ice mixture and in a N$_2$/CO layered ice at 16~K. In the CO$_2$ experiments the temperature is set to 16 -- 60~K, the irradiation flux to $1.1-8.3\times10^{13}$ photons cm$^{-2}$ s$^{-1}$ and the ice thickness to 2.1 -- 16.5~ML.

\begin{table*}
\begin{center}
\caption{Summary of CO and N$_2$ experiments}             
\label{con2exps}      
\centering                          
\begin{tabular}{l l c c c c c}        
\hline\hline                 
Experiment & Composition & Temperature (K)&Thickness (ML)&Lamp flux ($10^{13}$  photons cm$^{-2}$ s$^{-1}$)\\    
\hline                        
1&$^{13}$CO&15&4&4.7\\
2&$^{13}$CO&22&3.5&4.7\\
3&$^{13}$CO&27&3.5&4.7\\
4&$^{13}$CO$^{a \rm}$&16&3&4.7\\
5&$^{15}$N$_2$&16&4&4.7&\\
6&$^{13}$CO:$^{15}$N$_2$ mixed&16&4:4&4.7\\
7&$^{15}$N$_2$/$^{13}$CO layered&16&1/4&4.7\\
\hline                                   
\end{tabular}
\end{center}
$^{a \rm}$annealed ice (i.e. deposited at 27 K and then cooled down to 16 K)
\end{table*}

\begin{table*}
\begin{center}
\caption{Summary of CO$_2$ experiments}             
\label{co2exps}      
\centering                          
\begin{tabular}{l l c c c }        
\hline\hline                 
Experiment & Composition & Temperature (K)&Thickness (ML)&Lamp flux ($10^{13}$  photons cm$^{-2}$ s$^{-1}$)\\
\hline                        
1&$^{13}$C$^{18}$O$_2$&16&5.5&4.7\\
2&$^{13}$C$^{18}$O$_2$&18&2.1&2.3\\
3&$^{13}$C$^{18}$O$_2$&18&5.5&2.3\\
4&$^{13}$C$^{18}$O$_2$&18&5.6&2.3\\
5&$^{13}$C$^{18}$O$_2$&18&9.0&2.3\\
6&$^{13}$C$^{18}$O$_2$&18&16.5&2.3\\
7&$^{13}$C$^{18}$O$_2$&18&3.9&1.1\\
8&$^{13}$C$^{18}$O$_2$&18&4.7&3.5\\
9&$^{13}$C$^{18}$O$_2$&20&6.2&8.3\\
10&$^{13}$C$^{18}$O$_2$&18&6.5&8.3\\
11&$^{13}$C$^{18}$O$_2$ $^{a \rm}$&18&7.0&2.3\\
12&$^{13}$C$^{18}$O$_2$ $^{a \rm}$&16&4&4.7\\
13&$^{13}$C$^{18}$O$_2$&30&6.2&2.3\\
14&$^{13}$C$^{18}$O$_2$&40&5.8&2.3\\
15&$^{13}$C$^{18}$O$_2$&50&6.7&2.3\\
16&$^{13}$C$^{18}$O$_2$&60&3.3&2.3\\
17&$^{13}$C$^{18}$O$_2$&60&7.4&2.3\\
18&$^{13}$C$^{18}$O$_2$&60&5.8&1.1\\
19&$^{13}$C$^{18}$O$_2$&60&6.2&8.3\\
20&$^{13}$C$^{18}$O$_2$&60&11.0&2.3\\
21&$^{13}$CO$_2$&16&3.8&4.7\\
22&$^{13}$CO/$^{13}$C$^{18}$O$_2$ $^{b \rm}$&18&10/5&2.3\\
23&N$_2$/$^{13}$C$^{18}$O$_2$ $^{b \rm}$&18&20/5.4&2.3\\
24&$^{13}$C$^{18}$O$_2$/CO$_2$ $^{b \rm}$&18&6.5/7&2.3\\
25&$^{13}$C$^{18}$O$_2$/H$_2$O $^{b \rm}$&18&7/10&2.3\\
\hline                                   
\end{tabular}
\end{center}
$^{a \rm}$annealed ice (i.e. deposited at 60 K and then cooled down to 16 or 18 K)
$^{b \rm}$layered ices
\end{table*}

\subsection{Data analysis}

\begin{figure}
\resizebox{\hsize}{!}{\includegraphics{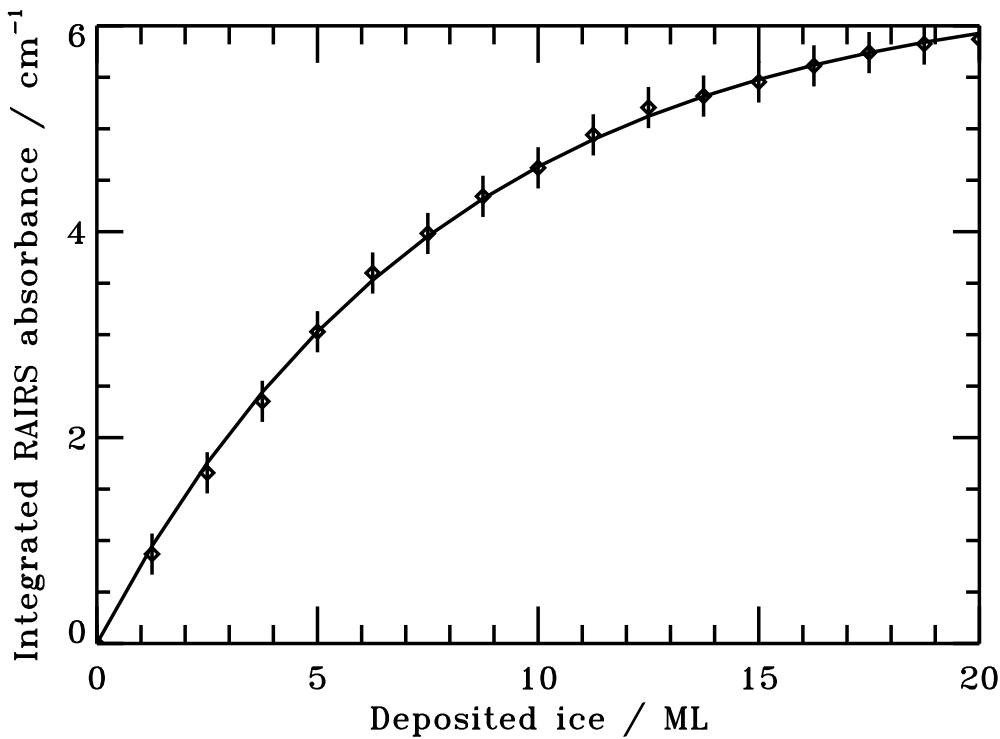}}
\caption{The integrated absorbance of the $^{13}$C$^{18}$O$_2$ stretching band as a function of deposited ice. The fitted exponential function is used to correct for non-linear growth of the integrated absorbance above 5~ML. Below 5 ML the RAIRS absorbance and the ice thickness are linearly correlated within the experimental uncertainties.}
\label{rairs_abs}
\end{figure}

\begin{figure}
\resizebox{\hsize}{!}{\includegraphics{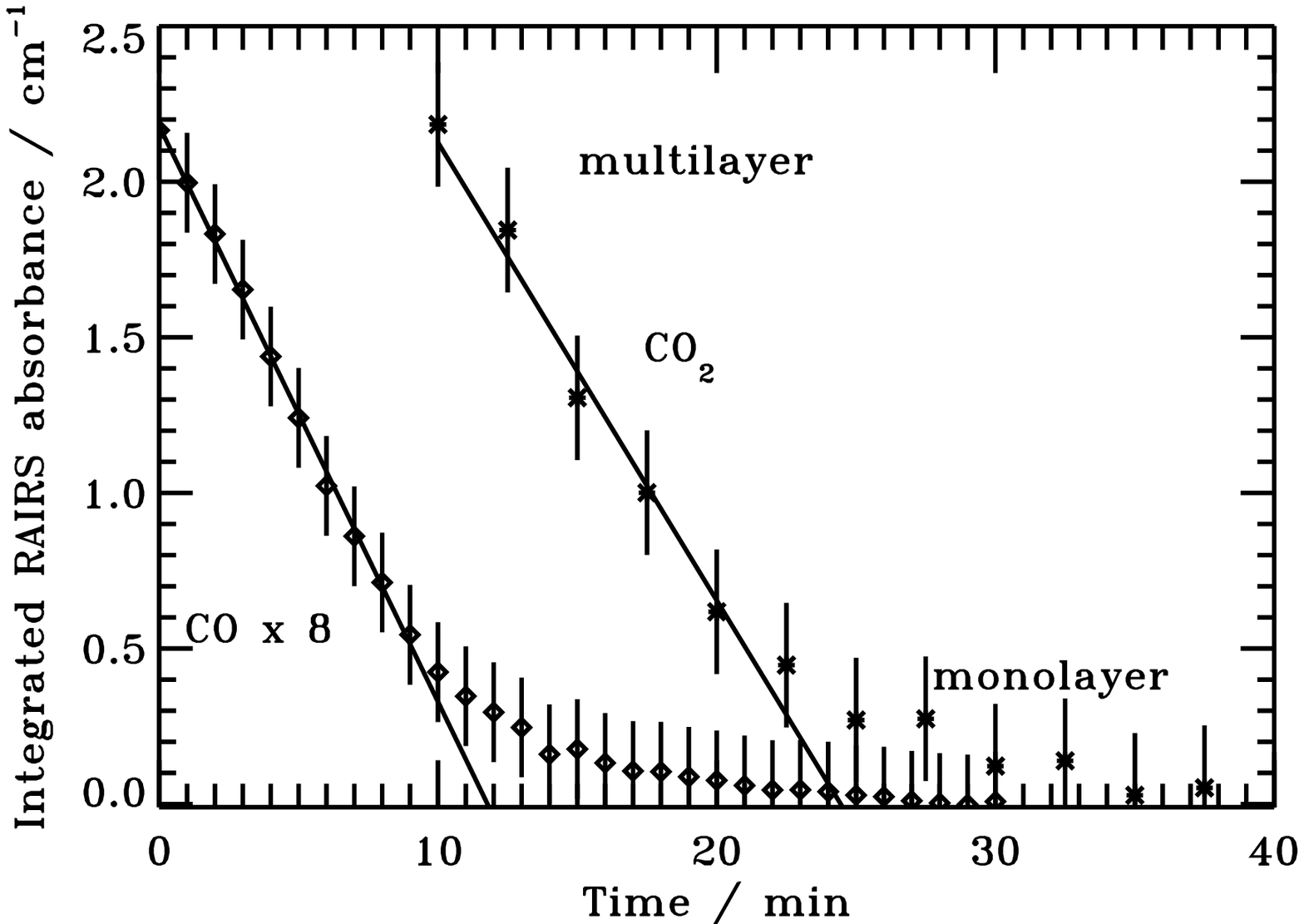}}
\caption{The integrated absorbance of the $^{13}$C$^{18}$O (diamonds) and $^{13}$C$^{18}$O$_2$ (crosses) stretching bands as a function of time during isothermal desorption of $\sim$ 5~ML ices at 31 and 76~K, respectively. The full lines are used to distinguish between the constant desorption rate in the multilayer regime and the decreasing desorption rate once the monolayer regime is reached.}
\label{abs_ml}
\end{figure}

The UV induced ice loss yield during each CO and CO$_2$ experiment is determined by RAIRS of the ice during irradiation. The intensity of the RAIRS profile is linearly correlated with the ice layer thickness of CO  and CO$_2$ ice up to $\sim$5~ML, but the RAIRS profile can be used up to 20~ML for analysis as long as the non-linear growth above 5~ML is taken into account  (Fig. \ref{rairs_abs}). The absolute loss yield in number of molecules lost per incident photon is calculated from the RAIRS intensity loss as a function of UV fluence.

For this calculation it is vital to have good estimates of the CO and CO$_2$ band strengths. Due to the fact that all ice measurements are done using RAIRS, the ice thickness cannot be estimated from previously determined ice transmission band strengths. Instead the band strength of one ice monolayer is calculated from the observed difference in isothermal desorption from multilayer coverages (constant rate) and monolayer coverages (decreasing rate) as shown in Fig. \ref{abs_ml}. For CO this was done at 31~K and for CO$_2$ at 76~K. The integrated absorbance of 1~ML is estimated to within 40\% from the RAIR spectra at this turning point, which in its turn is used to calculate a band strength relevant for RAIRS. The calculated band strengths are 0.07 and 0.55 cm$^{-1}$ ML$^{-1}$ for CO and CO$_2$ at their respective desorption temperatures. At 18 K the bands are somewhat weaker at 0.06 and 0.42 cm$^{-1}$ ML$^{-1}$ , respectively. These values are highly set-up specific and they depend on such experimental parameters as mirror settings and should not be used for other purposes. This technique is based on the assumption that the ice is quite flat at the desorption temperature, which was confirmed by the previous CO experiments \citep{Oberg07b}. For CO, the deduced ice thicknesses agree well (within 20\%) with the theoretical values for our chosen deposition pressure and deposition time \citep{adsorption}. For CO$_2$, the measured ice thickness is $\sim$30\% lower than predicted, probably due to the fact that some of the CO$_2$ freezes out on the heating shield rather than depositing onto the substrate. Using this method we find that the relative band strengths of CO and CO$_2$ ice compare well (within 20\%) with previously measured transmission band strengths \citep{Hudgins93, Gerakines96}. To convert the band strengths from cm$^{-1}$ ML$^{-1}$ to cm molecule$^{-1}$ a surface density of  $\sim$10$^{15}$ molecules cm$^{-2}$ is assumed.

In the case of CO, there is no measurable photodissociation in this wavelength range and the measured photon-induced loss yield is the photodesorption yield \citep{Gerakines96, Cottin03, Oberg07b}. Simultaneous QMS measurements of the desorbed CO gas phase molecules allow for the calibration of the QMS signal to an absolute photodesorption yield. The calibrated QMS signal for CO is used to determine the fraction of the CO$_2$ ice that photodesorbs as CO. It is also used, together with the measured relative sensitivities of the QMS filament to CO and N$_2$, to determine the N$_2$ photodesorption yield. The N$_2$ photodesorption yield cannot be determined by RAIRS since N$_2$ has no permanent dipole moment and thus no strong IR feature.

In contrast to CO, CO$_2$ has only dissociative transitions in the wavelength region of the lamp; a UV photon absorption dissociates CO$_2$ into CO + O with a quantum yield of up to 98\% in the gas phase \citep{Slanger78}. Hence, UV irradiation induces chemistry as well as desorption \citep{Gerakines96}. To determine the total photodesorption yield, the ice loss due to desorption must be separated from the conversion of CO$_2$ into other ice products. This is done by analysis of the RAIR spectra using two different methods: kinetic modeling and mass balance. The first method uses the different kinetics of surface processes, like desorption, and bulk processes, like ice photolysis, to distinguish between the two. Surface desorption from a multilayer ice is expected to be a zeroth order process and the photodesorption yield, which is determined from the derivative of the ice thickness with fluence, should therefore be constant with UV fluence. Photolysis into other species occurs throughout the ice at equal yield, since the ices here are thin enough that optical depth effects can be ignored, and is consequently expected to be a first order process. The contributions of desorption and photolysis to the observed ice loss is then determined by fitting the observed ice thickness versus fluence to the sum of a linear function and an exponential decay function, corresponding to a zeroth and a first order reaction. This method has the advantage that it is not dependent on identifying the photolysis products of CO$_2$ and it is used to derive photodesorption yields whenever the zeroth and first order curves are separable. This is mainly the case for the high temperature and high fluence experiments.

The mass balance analysis method compares the total ice loss with the simultaneous formation of other species in the ice; the final photodesorption yield is then defined as the loss yield of the original ice minus the formation yield of other carbon-bearing ice species. This is the only method that works for ices that are exposed to low fluences, where the contributions from the zeroth and first order processes cannot be separated. In the CO$_2$ experiments CO, CO$_3$ and O$_3$ are the expected reaction products \citep{Gerakines96}, with CO dominating. The photodesorption yield is then the CO$_2$ loss yield minus the formation yield of CO and CO$_3$, though as seen below the contribution from CO$_3$ is negligible. This method depends on the relative infrared band strengths of the formed molecules and is thus only accurate in the temperature range where the CO band strengths are measured, i.e. $<$30~K. As seen below, these two methods agree very well in the few cases where both methods are used to derive photodesorption yields.

The simultaneous mass spectrometry of gas phase molecules during irradiation reveals the nature of the desorbed species. This is limited by the fact that less volatile molecules (e.g. CO$_2$) adsorb onto the heating shield and other semi-cold surfaces inside the experiment before reaching the mass spectrometer. In the case of CO$_2$, only the fractions of the ice that desorb as CO and other volatile species are detected by the QMS. At temperatures above 30~K the conversion factor between QMS detected and desorbed CO changes due to a decrease in cryopumping of CO, which is accounted for when deriving the CO-from-CO$_2$ photodesorption yield. For both CO and CO$_2$, re-condensation onto the actual ice sample after desorption will play a negligible role given the small surface area of the sample and the resulting underestimate of the actual photodesorption will be substantially lower than other sources of inaccuracy.

To summarize, the main sources of uncertainty in these experiments are the photon flux and ice thickness calibrations of $\sim$30\% and 40\%, respectively. In addition, from repeated experiments, the CO and CO$_2$ experimental results are found to vary with approximately 10\% and 25\%, respectively. The uncertainty is greater for CO$_2$ than for CO because of the extra steps in deriving the CO$_2$ photodesorption yield. The total uncertainty is $\sim$60\% for the CO$_2$ photodesorption yield and $\sim$50\% for the CO photodesorption yield. The N$_2$ photodesorption yield uncertainty is somewhat higher due to the larger uncertainty in the QMS measurements. 

\section{Results}

\subsection{CO and N$_2$}

\begin{figure}
\resizebox{\hsize}{!}{\includegraphics{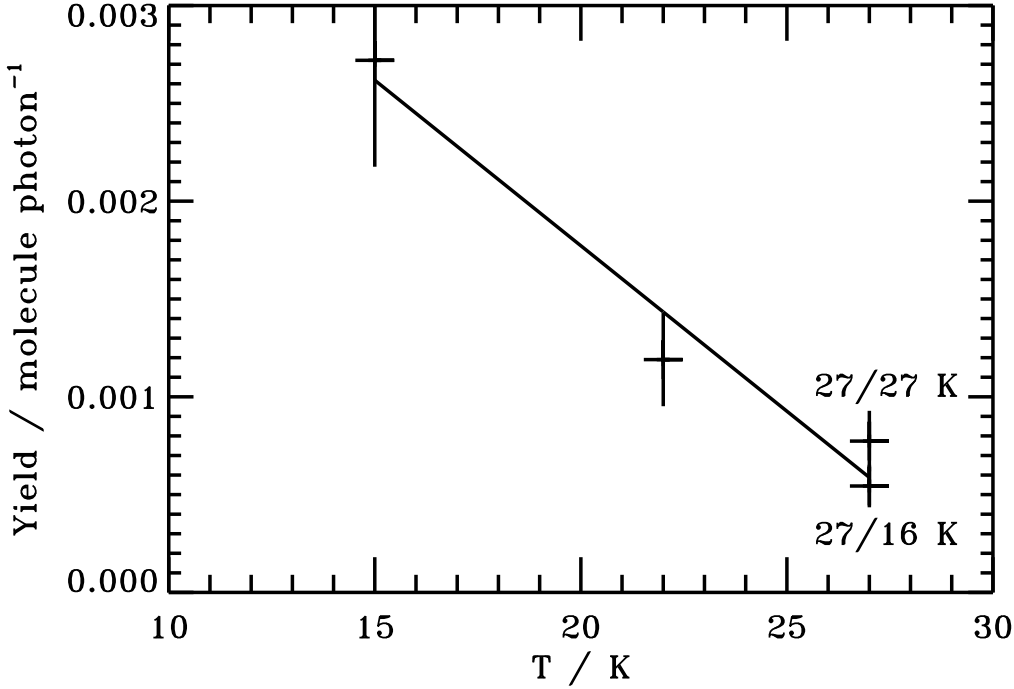}}
\caption{The CO ice photodesorption yield as a function of ice deposition temperature between 15 and 27~K. The ices are photodesorbed at the deposition temperature except for the point marked 27/16~K, where the ice is deposited at 27~K and then cooled down and irradiated at 16~K. The plotted uncertainties are the relative uncertainties between different experiments. The uncertainty in the absolute photodesorption yield is 50\%.}
\label{coTrates}
\end{figure}

\begin{figure}
\resizebox{\hsize}{!}{\includegraphics{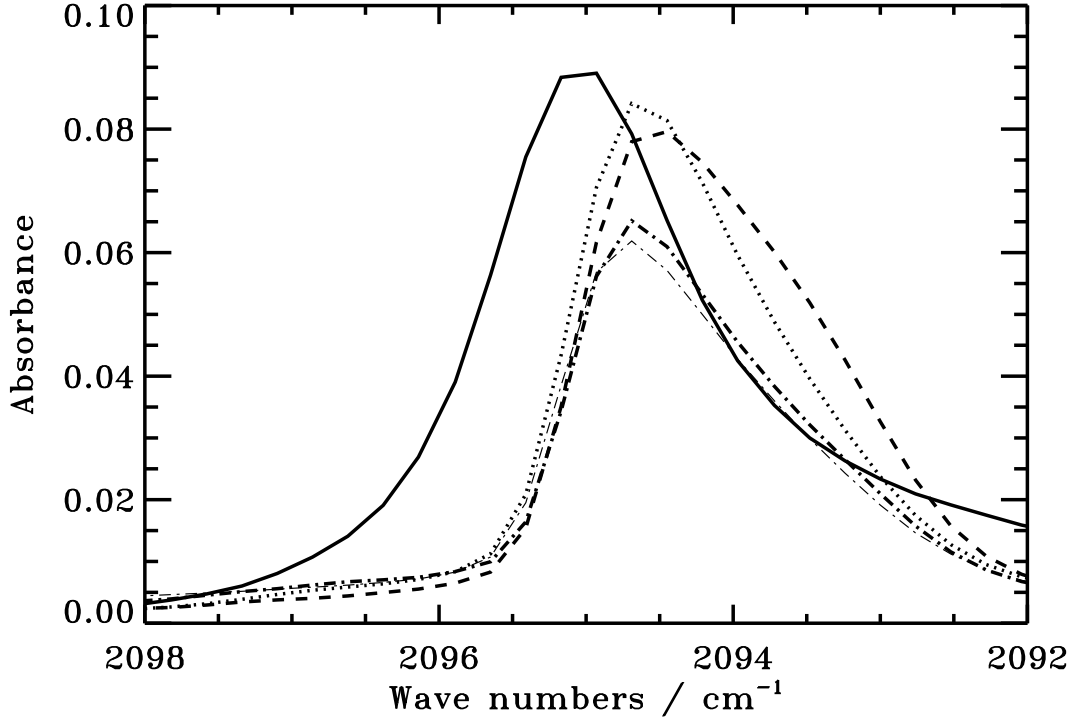}}
\caption{CO spectra at 15~K (solid), 22~K (dotted), 27~K (dashed) and at 16~K and at 16 K of a sample deposited at 27~K (dash dotted), all acquired before irradiation. The figure also shows a spectrum of the annealed ice acquired after a UV fluence of $\sim7\times10^{17}$ photons cm$^{-2}$ (thin dash dotted).}
\label{co_sp}
\end{figure}

\begin{figure}
\resizebox{\hsize}{!}{\includegraphics{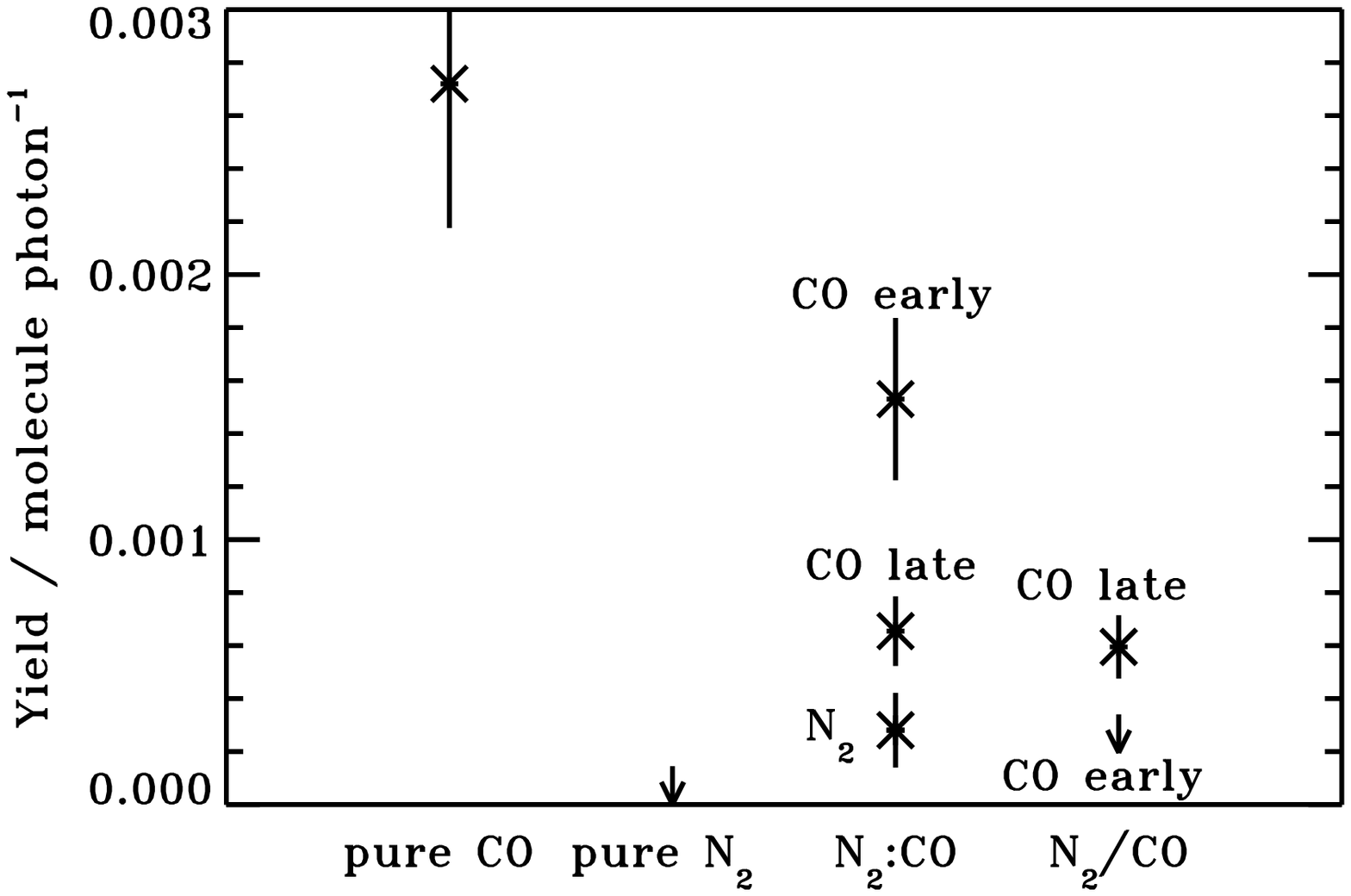}}
\caption{The CO and N$_2$ desorption yields and upper limits in 4~ML pure CO and pure N$_2$ ices, a 4:4~ML mixed ice, and a N$_2$/CO 1/4~ML layered ice, all at 16~K, except for the pure CO ice at 15~K. In the mixed and layered ices the CO desorption yields are not constant and 'early' and 'late' marks the CO desorption yield in the beginning of the experiment and after a photon fluence of $8.5\times10^{17}$ cm$^{-2}$.}
\label{con2}
\end{figure}

The results from photodesorption experiments of pure CO ice at 15~K are reported in \citet{Oberg07b}.  The yield presented there is updated here using new ice thickness and lamp flux calibrations; at 15~K the CO photodesorption yield is (2.7$\pm$1.3) $\times 10^{-3}$ CO molecules per \textit{incident} photon, averaged over the wavelength range of the lamp. The corresponding photodesorption quantum efficiency per \textit{absorbed} UV photon in the surface layer is estimated using the lamp spectrum from \citet{Munozcaro03} and the UV spectrum of CO ice from \citet{Mason06}. The UV ice absorption spectrum shows that the CO lines are resolved. The measured absorption spectrum $\int^{\rm UV} A^{\rm ice}_{\rm \lambda} {\rm d}\lambda$ is not calibrated to a UV cross section so in our calculation it is assumed that the total UV absorption cross section is the same in the ice as measured in the gas phase $\int^{\rm UV} \sigma^{\rm gas}_{\rm \lambda} d\lambda$ \citep{Eidelsberg92, Eidelsberg99}.  The fraction of incident photons that are absorbed in the top monolayer, $\chi_{\rm abs}$, is calculated from 

\begin{equation}
\chi_{\rm abs} = \frac{\int^{\rm UV}I^{\rm UV-lamp}_{\rm \lambda}A^{\rm ice}_{\rm \lambda}{\rm d}\lambda}{\int^{\rm UV} I^{\rm UV-lamp}_{\rm \lambda} {\rm d}\lambda} \times \frac{\int^{\rm UV} \sigma^{\rm gas}_{\rm \lambda} {\rm d}\lambda}{\int^{\rm UV} A^{\rm ice}_{\rm \lambda} {\rm d}\lambda} \times N_{\rm s}\\
\label{crosscal}
\end{equation}

\noindent by cross-correlating the UV-lamp spectrum with the absorption spectrum of CO ice, divided by the total UV flux, $\int^{\rm UV} I^{\rm UV-lamp}_{\rm \lambda} {\rm d}\lambda$, and then multiplying with the cross section conversion factor and the amount of molecules in one monolayer, $N_{\rm s}$. The resulting absorption fraction is $5.5\pm0.2\times 10^{-3}$, where the uncertainty reflects the error in the gas phase cross section. Comparison with our measured photodesorption yield results in an efficiency of 0.3--0.8 per absorbed photon at 15~K, including both the absorbance and the photodesorption uncertainties.

Figure \ref{coTrates} shows that the CO photodesorption yield decreases with ice temperature such that it is almost a factor of three lower at 27~K compared to 15~K. Within this temperature range the CO photodesorption yield is empirically described as linearly dependent on temperature: $2.7\times10^{-3}-(T-15)\times1.7\times10^{-4}$ molecules photon$^{-1}$, where $T$ is temperature in K. An additional experiment, where the ice is deposited at 27~K and then cooled down to 16~K before irradiation, results in a similar desorption yield as when the ice is also desorbed at 27~K. In quantum efficiency terms this corresponds to 0.1--0.3 photodesorption events per absorbed photon in the top ice layer, at 27~K as well as for the annealed ice at 16~K. This indicates that the structure of the ice, rather than the temperature, affects the photodesorption yield. This is further supported by a change in RAIRS profile at 22 and 27~K compared to that at 15~K (Fig. \ref{co_sp}). Changes in the CO spectra with temperature have been previously reported by e.g. \citet{Fuchs06}. These spectral profiles do not change visibly after a UV fluence of $7\times10^{17}$ photons cm$^{-2}$ when the ices are irradiated at their deposition temperature (not shown). Figure \ref{co_sp} also shows the spectral profile of the annealed ice before and after irradiation, which has not changed significantly with cool down, and it has at most slightly shifted toward the 15~K ice spectral profile following irradiation.

In \cite{Oberg07b} the N$_2$ photodesorption yield is constrained to a factor of 10 less than the CO yield at 15~K. With increased sensitivity of the mass spectrometer, N$_2$ photodesorption is now detected at a yield of $1.8\times10^{-4}$ N$_2$ molecules photon$^{-1}$ or a factor of 15 lower than the CO photodesorption yield. This value has a factor of two uncertainty, mainly due to the uncertainty in the conversion of the mass spectrometer signal into an ice desorption yield. This measured yield is real, but because of continuous freeze-out of $\sim$0.1 ML H$_2$O ice per hour the N$_2$ ice contains an H$_2$O impurity. A typical experiment lasts 5--6 hours resulting in a maximum 12\% contamination level (a significant fraction of the adsorbed H$_2$O molecules photodesorbs themselves during the irradiation experiments). This probably affects the measured photodesorption yield due to co-desorption of ices and thus the measured yield should be used as a strict upper limit of pure N$_2$ photodesorption. 

In two additional experiments a CO:N$_2$ mixture 4:4~ML and a N$_2$/CO layered 1/4~ML ice are irradiated at 16~K. Figure \ref{con2} shows that in the ice mixture experiment the N$_2$ desorption yield more than doubles compared to pure N$_2$ ice, while the CO desorption yield decreases by a factor of 2--4 compared to pure CO ice during the experiment. After a fluence of $\sim2\times10^{17}$ photons cm$^{-2}$ the CO desorption yield is 50\% of the photodesorption yield of pure CO. This is expected in a 1:1 mixture, since only 50\% of the surface is covered with CO. With increasing fluence the CO desorption yield decreases such that it is only 25\% of the yield of pure CO photodesorption after $8.5\times10^{17}$ photons cm$^{-2}$. This can only be understood if the N$_2$ molecules desorb with a lower yield than the CO molecules, resulting in a decreasing CO surface concentration with UV fluence. In the layered experiment the CO photodesorption yield is initially below the detection limit. After a UV fluence of $8.5\times10^{17}$ photons cm$^{-2}$ the yield increases to 25\% of the pure CO photodesorption yield. The N$_2$ mass spectrometry signal does not reach equilibrium in the layered experiment, but the desorption yield seems to be at a similar level as in the mixed ice.

\subsection{CO$_2$}

\subsubsection{Derivation of the total photodesorption yield}

\begin{figure}
\resizebox{\hsize}{!}{\includegraphics{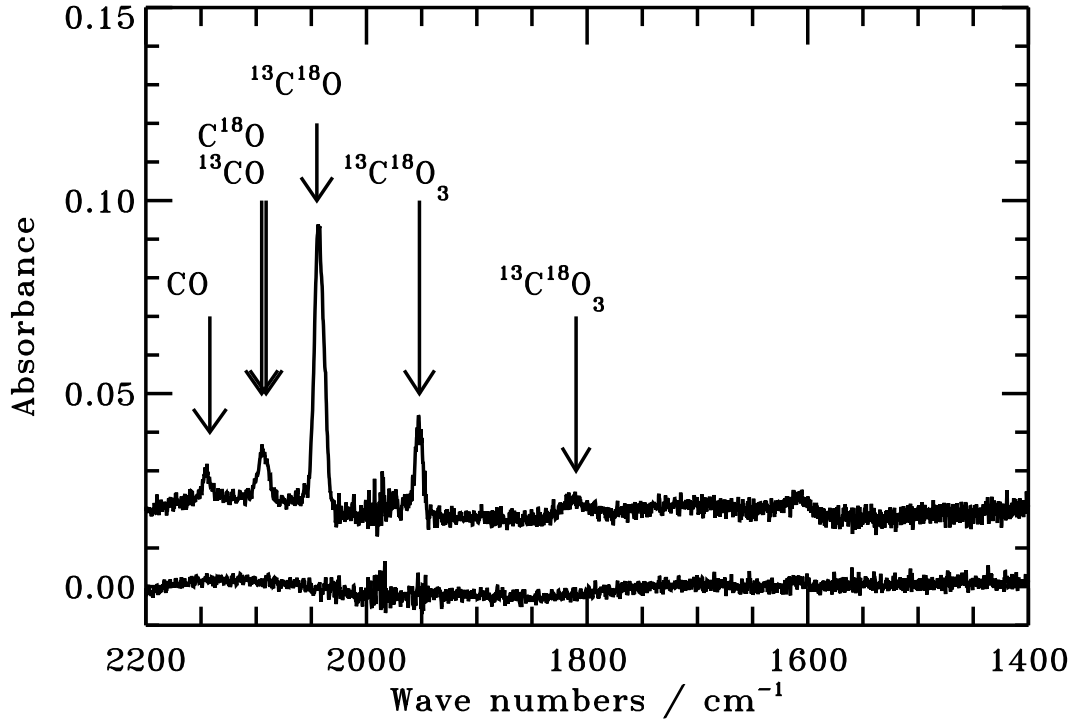}}
\caption{$^{13}$C$^{18}$O$_2$ ice at 18~K before (bottom) and after (top) a UV fluence of $5\times10^{17}$ photons cm$^{-2}$. Some of the original CO$_2$ ice is photolyzed into CO, CO$_3$ ($\nu_1$ at 1953~cm$^{-1}$ and 2$\nu_4$ in Fermi resonance with $\nu_1$ at 1810 cm$^{-1}$).}
\label{full_sp}
\end{figure}

\begin{figure}
\resizebox{\hsize}{!}{\includegraphics{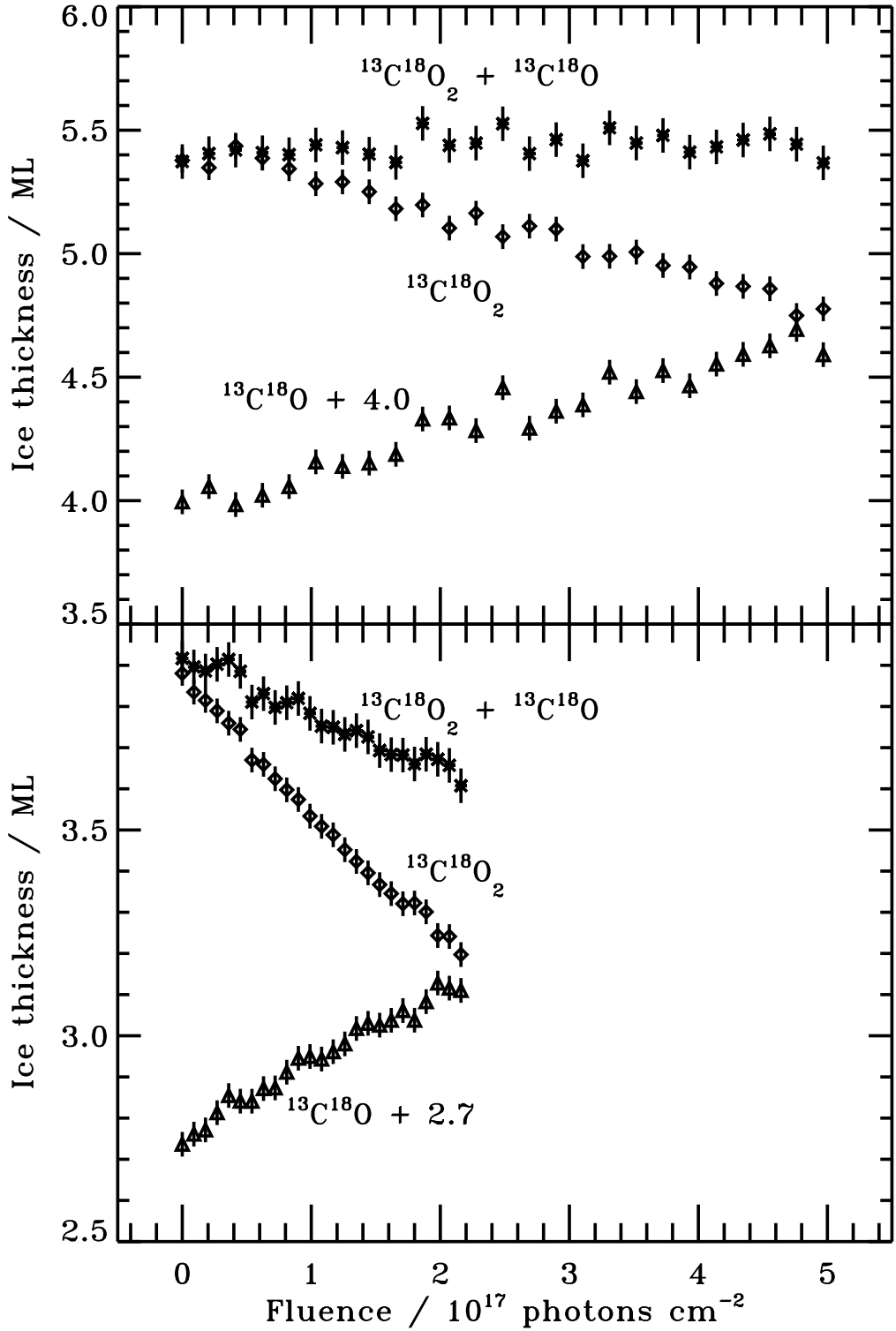}}
\caption{Top: the derived $^{13}$C$^{18}$O$_2$ ice thickness (diamonds)  in a 18~K, 20/5.4~ML N$_2$/$^{13}$C$^{18}$O$_2$ layered ice as a function of fluence, plotted together with the formed $^{13}$C$^{18}$O ice (triangles) and the calculated total ice thickness (stars). The latter is constant with fluence within the experimental uncertainties. The bottom panel shows the decreasing total ice thickness with UV fluence in an 18~K, 3.9 ML, uncovered experiment. In these plots the error bars indicate the relative uncertainty in the integrated absorbance (converted to a ML scale) of the RAIRS features within each experiment. This is also the case for similar plots throughout the paper}
\label{des_test}
\end{figure}

\begin{figure}
\resizebox{\hsize}{!}{\includegraphics{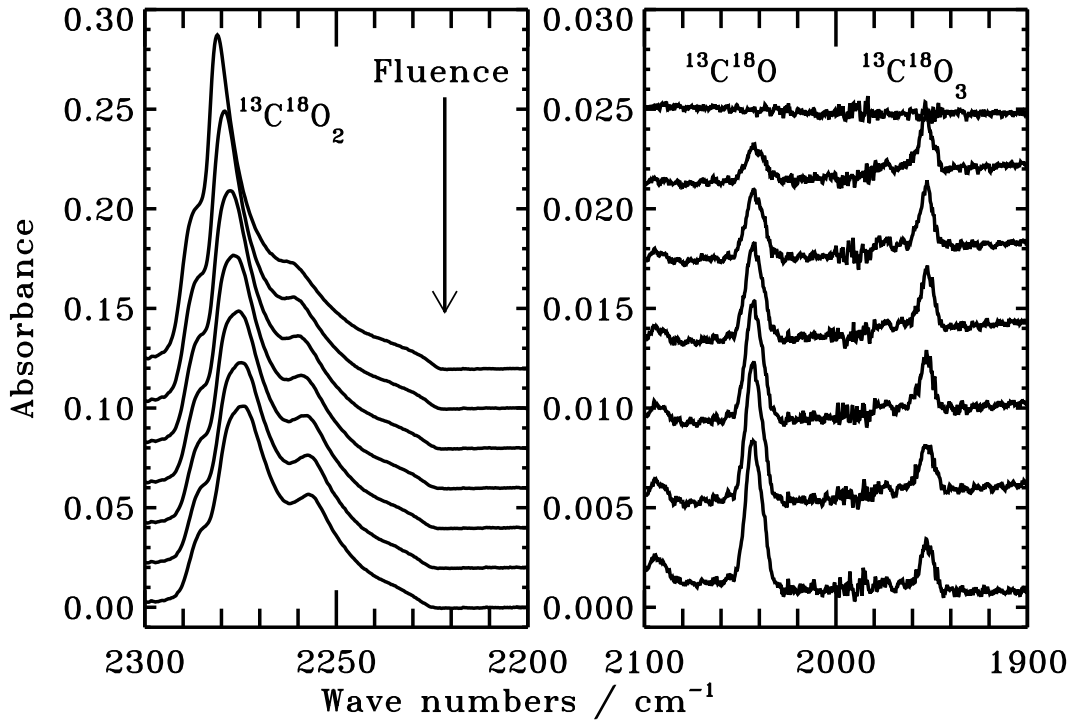}}
\caption{RAIR spectra of the $^{13}$C$^{18}$O$_2$ stretching band at 2280 cm$^{-1}$, the $^{13}$C$^{18}$O stretching band at 2045~cm$^{-1}$ and the $^{13}$C$^{18}$O$_3$ $\nu_1$ band at 1950~cm$^{-1}$ acquired before irradiation of a 11~ML $^{13}$C$^{18}$O$_2$ ice at 18~K and then after every $8.3\times10^{16}$ photons cm$^{-2}$. The absorbance of the CO$_2$ band decreases with UV fluence due to photodesorption and photodissociation, while the CO band absorbance increases. Note the nearly constant $^{13}$C$^{18}$O$_3$ integrated absorbance after a fluence of $\sim$8$\times10^{16}$ photons cm$^{-2}$.} 
\label{co2sp}
\end{figure}

\begin{figure}
\resizebox{\hsize}{!}{\includegraphics{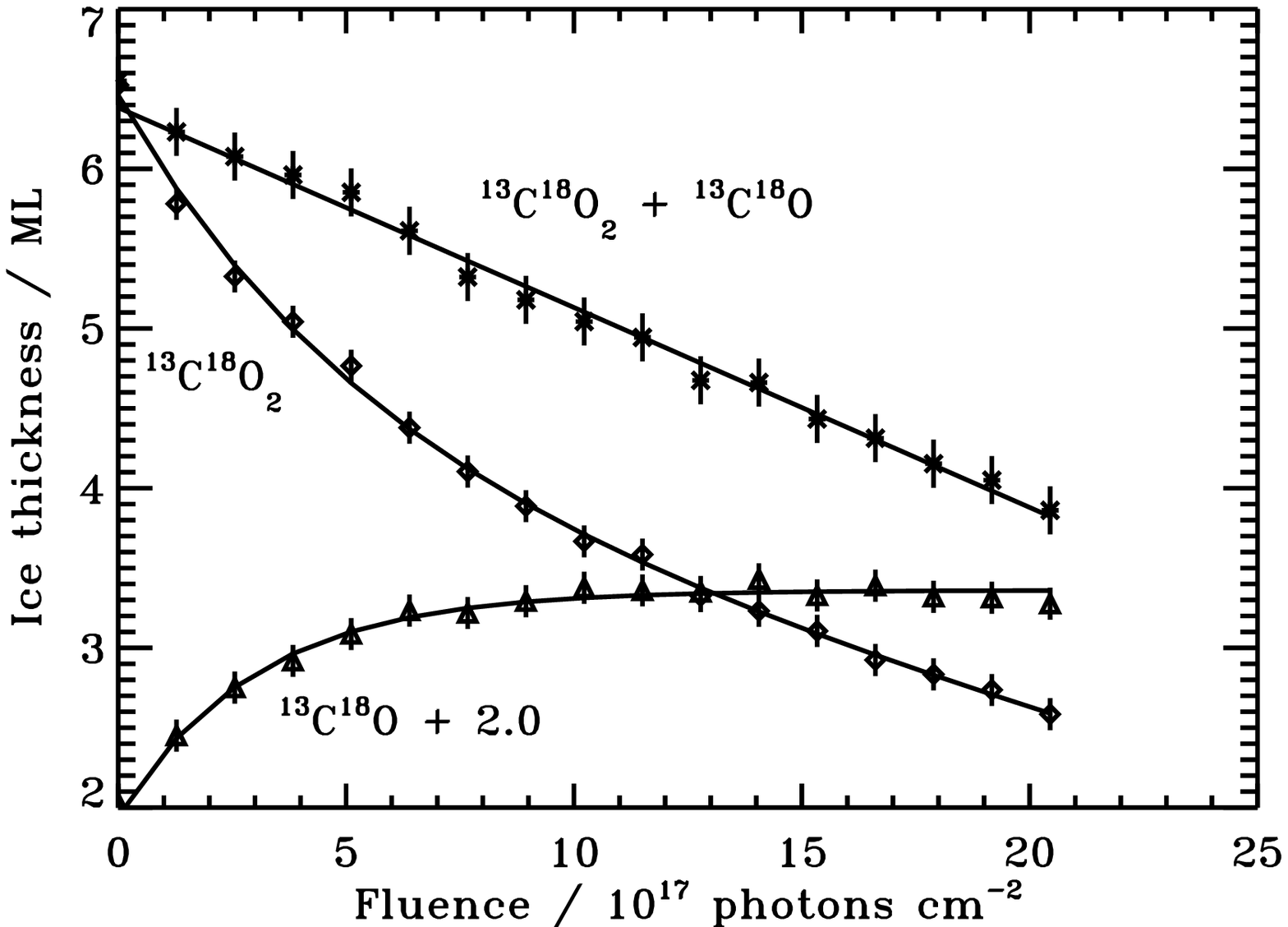}}
\caption{The calculated layer thicknesses of an originally 6.5~ML CO$_2$ ice at 18~K as a function of UV fluence (diamonds) plotted together with the formed CO layer thickness (triangles) and the calculated total ice thickness (stars). The CO$_2$ ice loss is fitted as a sum of photolysis (an exponential function) and desorption (a linear function). The exponential part mirrors the CO formation within the fit uncertainties.}
\label{co2rate}
\end{figure}

To use the mass balance method to calculate the CO$_2$ photodesorption yield, it is necessary to constrain which molecules are formed during irradiation. Figure \ref{full_sp} shows the spectra of an 18~K, 9~ML thick CO$_2$ ice before and after a UV fluence of $5\times10^{17}$ photons cm$^{-2}$. The only formed molecules are CO and CO$_3$, though O$_3$ formation cannot be excluded since the strong $\nu_3$ $^{18}$O$_3$ band around 1040~cm$^{-1}$ is outside of the range of the detector. This is in agreement with \citet{Gerakines96}, who found CO, CO$_3$ and small amounts of O$_3$ after irradiating CO$_2$ with a similar fluence. The line positions are taken from \citet{Brewer72}, \citet{Moll65} and \citet{Gerakines01}. The weak band around 1605 cm$^{-1}$ cannot be unambiguously assigned. The lack of other features in the spectra from e.g. carbon-suboxides put strict upper limits on the formation of such molecules to a fraction of a percentage of the original CO$_2$ ice. This is consistent with the experiments of \citet{Gerakines01} where no carbon-suboxides was detected after UV irradiation of pure CO$_2$ ice. In addition Temperature Programmed Desorption (TPD) experiments following irradiation show that $\sim$10\% of the original ice is photolyzed into CO, $\sim$1\% into O$_2$ or O$_3$ and $\sim$0.1\% into C$_2$. From this we infer that more than 99\% of the carbon budget is bound up in CO$_2$, CO and CO$_3$ during the experiment.

The CO$_2$ and CO abundances during each experiment are calculated from their derived band strengths. The CO$_3$ band strengths have not been measured, however, and can only be crudely estimated. This will introduce a large uncertainty into the mass balance calculations if CO$_3$ is formed at significant abundances. Figure \ref{des_test} shows the calculated CO$_2$ and CO ice thicknesses as a function of fluence for a layered 20/5.4~ML N$_2$/CO$_2$ ice. The ice cover hinders desorption and the result is that 10\% of the original CO$_2$ is photolyzed into CO. The fact that the lost CO$_2$ is perfectly compensated for by the formed CO ice shows that CO$_3$ is not a significant photolysis product. The lack of photodesorption in the layered experiment (top Fig. \ref{des_test}) is contrasted with the observed photodesorption of a 3.9 ML bare CO$_2$ ice (bottom Fig. \ref{des_test}).

The CO$_3$ abundance is also estimated independently by employing its only likely formation path and the fact that CO$_3$ reaches its final level within $\sim5\times10^{16}$ photons cm$^{-2}$ in all experiments (exemplified in Fig. \ref{co2sp} where the level does not change between 0.8 and 5$\times10^{17}$ photons cm$^{-2}$). CO$_3$ is expected to form from CO$_2$ + O, where the O originates from photolysis of another CO$_2$ molecule into CO + O. With photodesorption hindered, the amount of CO$_3$ then never exceeds the amount of CO in the ice. In the 20/5.4 N$_2$/CO$_2$ experiment, the CO$_3$ abundance reaches steady state when less than one percent of the CO$_2$ is converted into CO, which puts a 1\% upper limit on the formed CO$_3$. This is small compared to the ice loss during CO$_2$ photodesorption experiments, where typically 20\% of the ice is lost.

From these results the mass balance photodesorption yield is defined as the CO$_2$ ice loss yield minus the CO formation yield. Figure \ref{co2rate} shows the CO$_2$, CO and CO$_2+$CO ice thicknesses in an originally 6.5~ML thick CO$_2$ ice at 18~K as a function of UV fluence. Practically the photodesorption yield is derived from the slope of the total ice thickness as a function of fluence. This mass balance method of determining the photodesorption yield agrees very well with the yield determined through simultaneous kinetic modeling of bulk photolysis and photodesorption. In Figure \ref{co2rate} the CO$_2$ ice thickness is fitted to a function of the form $A(0)\times e^{-A(1)/\Phi}+A(2)+A(3)\times\Phi$ using the IDL script MPFIT, where $\Phi$ is the fluence. The photodesorption yield is determined from $A(3)$. The derived photodesorption yield is the same, within the uncertainties, to the yield derived from fitting a linear function to the CO$_2+$CO ice thickness. This confirms the validity of both methods.

Whichever method is used, the result is a linear coefficient, which gives a photodesorption yield in loss of ice monolayers per $10^{17}$ UV photons for a 6.5 ML ice at 18~K: ${0.27  \: \rm (ML)}/{\rm 2.1\: (10^{17} \: photons \: cm^{-2})}=0.13$ ML / ($10^{17}$ UV photons cm$^{-2}$). This is further converted into a photodesorption yield in CO$_2$ molecules per UV photon (7-10.5~eV): $Y_{\rm pd}={\rm 0.13\times10^{-17} \:ML \:photon^{-1} cm^{2}}\times {\left(10^{15} \: \rm molecules \: cm^{-2}/1 \rm \: ML \right)} = \: 1.3\times10^{-3}$ molecules photon$^{-1}$.

\subsubsection{Desorption products}

\begin{figure}
\resizebox{\hsize}{!}{\includegraphics{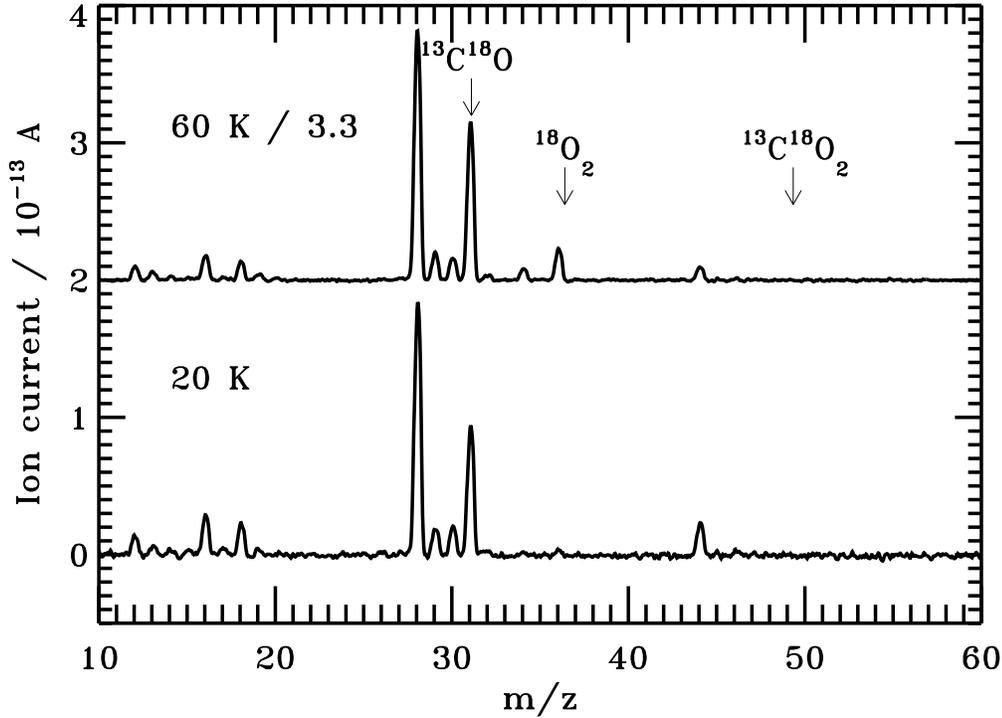}}
\caption{Mass spectra acquired during irradiation of a 6.2~ML thick $^{13}$C$^{18}$O$_2$ ice at 20 and 60~K  with a flux of 8.3$\times10^{13}$ photons cm$^{-2}$ s$^{-1}$. The spectra at 60~K have been divided by 3.3 to account for the lower cryopumping of volatiles like CO and O$_2$ at 60~K compared to at 20~K. In each case the ice is irradiated for 3 hours before acquisition to ensure that the photodesorption rate is stable. Each acquisition lasts 3 hours and consists of $\sim$100 averaged spectra. In addition to photodesorbed ices there are some background CO (m/z=12, 16 and 28), CO$_2$ (m/z=44) and possibly some background H$_2$O as well (m/z=18).}
\label{co2_ms}
\end{figure}

\begin{figure}
\resizebox{\hsize}{!}{\includegraphics{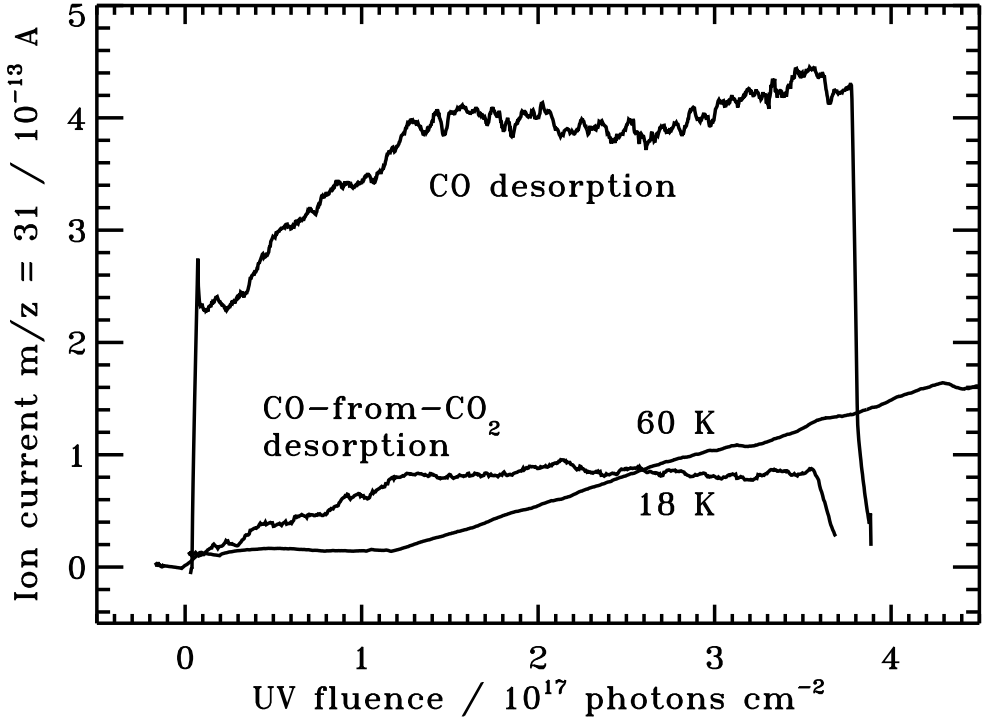}}
\caption{The detected CO photodesorption during irradiation of a 4 ML $^{13}$C$^{18}$O ice at 16~K and $^{13}$C$^{18}$O$_2$ ices at 18~K (5.5 ML) and at 60~K (7.4 ML), as a function of UV fluence. The 60~K signal has been divided by 3.3 to account for the lower cryo-pumping at 60~K compared to 18~K.}
\label{mscoco2}
\end{figure}

The total photodesorption yield is well constrained from the RAIR spectroscopy of the ice during irradiation. The question remains in which form CO$_2$ ice photodesorbs. Figure \ref{co2_ms} shows mass spectra acquired during UV irradiation of a 6.2~ML ice at the two temperature extremes, 20 and 60~K. The only visible desorption products are CO and O$_2$, which puts strict upper limits on other potential volatile desorption products. Less volatile species like CO$_2$ (m/z=49 for $^{13}$C$^{18}$O$_2$) cannot be excluded, however, since their cryopumping efficiencies are up to two orders of magnitude higher than for CO. As described in Section 2.2 the measured CO QMS signal can be converted into a number of CO molecules desorbed per photon. This number is compared with the total CO$_2$ photodesorption yield to quantify the amount of the CO$_2$ ice that desorbs as CO. From QMS measurements during irradiation experiments, $\sim20-50$\% desorbs as the fragment CO (Fig. \ref{mscoco2}) and at most 5\% as O$_2$. It is thus inferred that more than 50\% of the desorbed ice comes off as less volatile species, most likely CO$_2$. This is supported by the fact that the amount of formed CO$_3$ ice is the same whether or not the ice is covered and therefore whether or not photodesorption is allowed. This makes it unlikely that CO$_3$ is photodesorbing in the uncovered experiments. In addition no other C-bearing molecules are formed at significant abundances, which only leaves CO$_2$ as a possible desorption product.

Below we separate the total CO$_2$ ice photodesorption yield (as measured with RAIRS) from the desorption of CO-from-CO$_2$ (measured with the QMS). The desorption of CO$_2$ molecules is taken to be the difference between the total CO$_2$ ice desorption and the CO-from-CO$_2$ desorption. The total photodesorption quantum efficiency per absorbed UV photon in the surface layer is estimated to 0.4--1 using the lamp spectra from \citet{Munozcaro03} and the calibrated UV spectra of CO$_2$ ice from \citet{Mason06}. As seen from the thickness dependence below, this efficiency decreases with depth into the ice.

\subsubsection{Yield dependences on experimental parameters}

\begin{figure*}
\resizebox{\hsize}{!}{\includegraphics{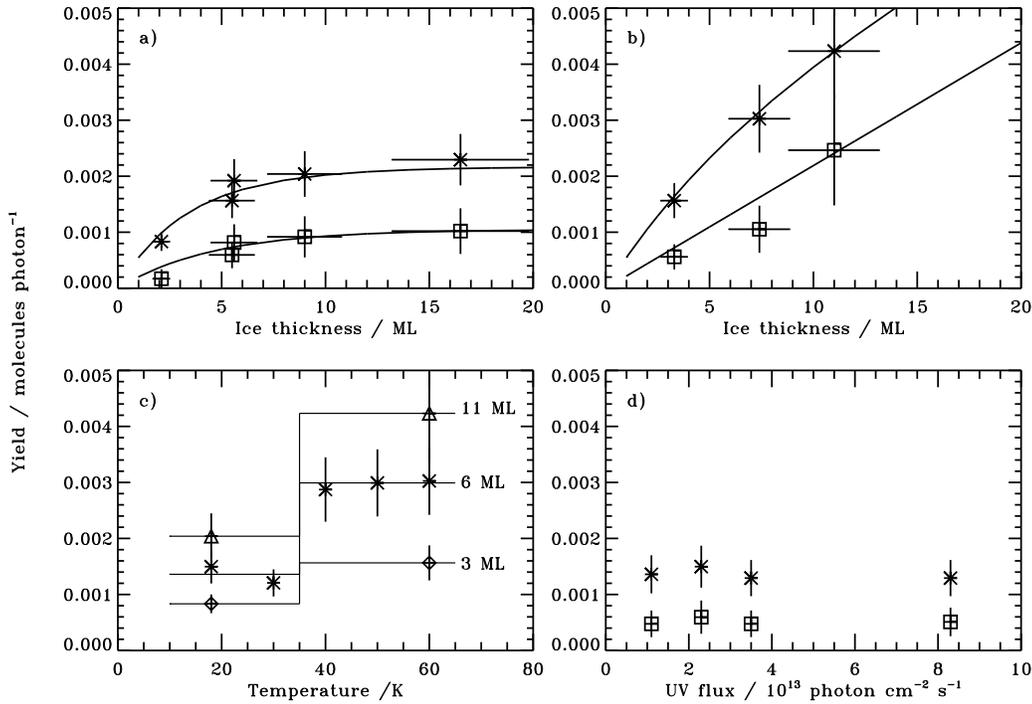}}
\caption{Total CO$_2$ (crosses) and CO-from-CO$_2$ (squares) photodesorption yield dependences on different parameters. In a) CO$_2$ ices of different thicknesses are irradiated with the same UV fluence of $\sim$6$\times10^{17}$ photons cm$^{-2}$ at 18~K and in b) at 60~K. Both the total CO$_2$ and the CO-from-CO$_2$ desorption yields are fitted with functions of the form $c \times (1-e^{-(x/l)} \rm$) at low temperatures (solid lines), where $x$ is the ice thickness and $l$ an ice diffusion parameter. The two measurements of 5.5 and 5.6~ML ices in a) are from the beginning and the end of a two-month long experimental series. Panel c) shows that the total photodesorption yield is constant with temperature within a low temperature region ($<$40~K) and within the warmer region 40--60~K for $\sim$3~ML (diamonds), 6~ML (crosses) and 11~ML (triangles) thick ices, irradiated with fluences of $\sim$6$\times10^{17}$ photons cm$^{-2}$. Finally panel d) demonstrates the independence of the total and CO-from-CO$_2$ photodesorption yields on the photon flux for 4--6.5~ML ices at 18~K. }
\label{combo}
\end{figure*}

\begin{figure}
\resizebox{\hsize}{!}{\includegraphics{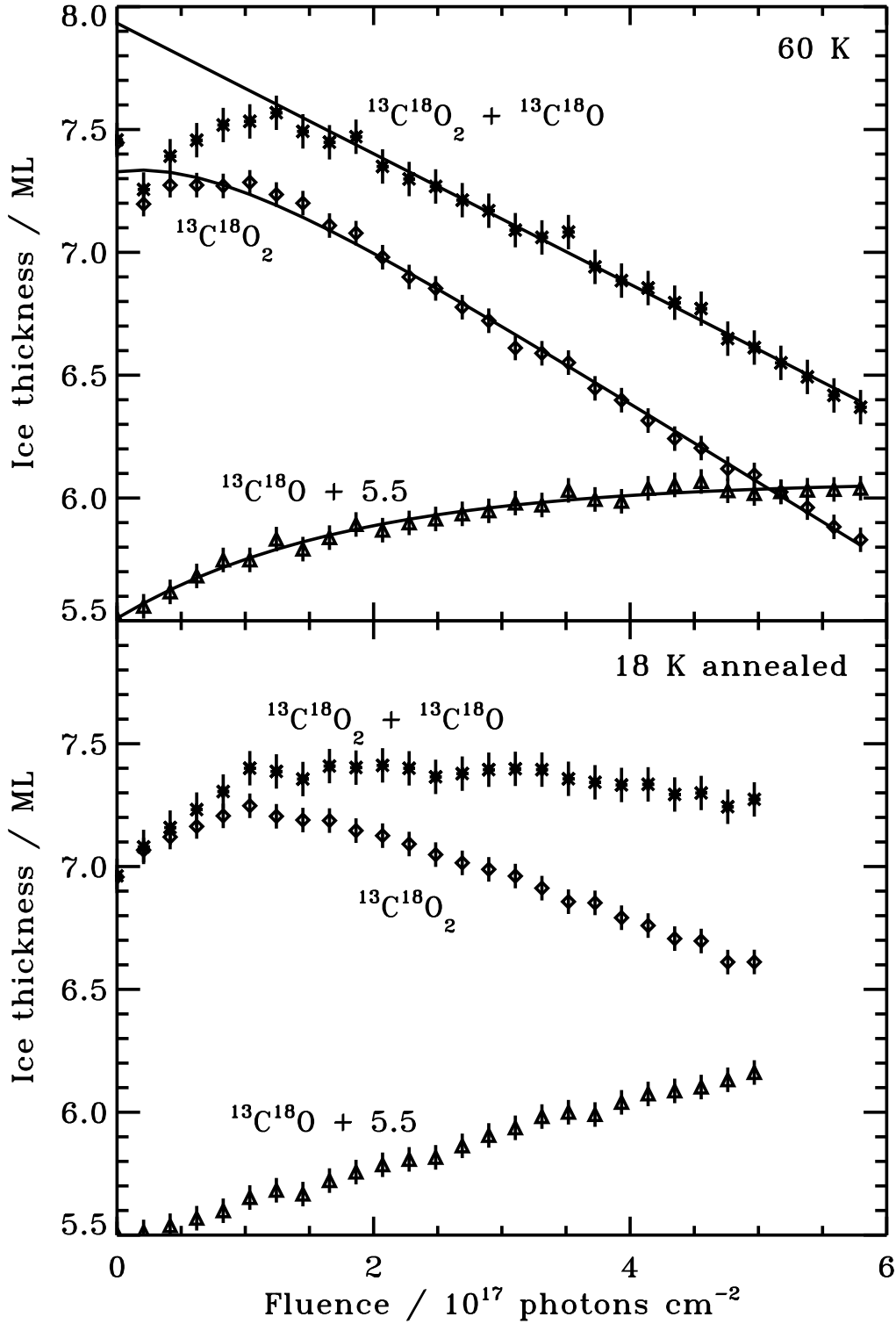}}
\caption{The top panel shows the measured  CO$_2$ layer thickness (diamonds) of an originally 7.4~ML CO$_2$ ice at 60~K as a function of UV fluence, plotted together with the formed CO layer thickness (triangles) and the calculated total ice thickness (stars). The CO$_2$ ice loss is fitted as a sum of photolysis (an exponential function) and desorption (a linear function). The bottom panel shows an 18~K experiment that has been thermally annealed at 60~K prior to irradiation at 18~K. Note the lack of evidence of photodesorption from this ice.}
\label{co2rate60k}
\end{figure}

\textit{Ice thickness.} The total CO$_2$ photodesorption yield at 18~K is thickness dependent up to several monolayers (Fig. \ref{combo}a), which is in contrast with the constant photodesorption yield of CO reported in \citet{Oberg07b}. The total CO$_2$ photodesorption yield increases from $8\times10^{-4}$ to $2.3\times10^{-3}$ molecules photon$^{-1}$ when the ice is grown from 2 to 16~ML. 

Figure \ref{combo}a also shows simple models fitted to the measured yields of both the CO-from-CO$_2$ and the total CO$_2$ photodesorption. The desorption yield of the CO-from-CO$_2$ is well described by $1.1(\pm0.2)\times10^{-3}\times(1- e^{-(x/4.6(\pm2.2) \rm)})$, where $x$ is the ice thickness in monolayers. The total CO$_2$ desorption dependence on thickness is a sum of the desorbed CO$_2$ and CO molecules and is thus modeled as $1.2(\pm0.1)\times10^{-3}\times (1-e^{-(x/2.9(\pm1.1) \rm)})+1.1(\pm0.2)\times10^{-3}\times(1- e^{-(x/4.6(\pm2.2) \rm)})$. Both models are fitted using the IDL routine MPFIT. Here the uncertainties only reflect the calculated model errors. As discussed in Section 2.2, the total uncertainty is 60\%. From these expressions it is clear that more than 90\% of the photodesorption events originate in the top 10 ML and 50\% in the top 3~ML of the ice. They also expose the thickness dependence of the fraction of the CO$_2$ desorbing as CO; between 2 and 16~ML this fraction grows from 20 to 45\%. The origin of the model and the full significance of the different exponential constants for CO and CO$_2$ desorption is further discussed below. 

The build-up of molecules in the 18~K CO$_2$ ice is linearly dependent on the ice thickness as expected for a photodesorption yield that is low in comparison to the total ice thickness. In all the ices, $\sim$10\% of the original ice is converted to CO and frozen into the CO$_2$ ice after a typical UV fluence of $5\times10^{17}$ photons cm$^{-2}$. For comparison $\sim$10\% of the original ice is desorbed after the same fluence in a typical 5 ML experiment.

Figure \ref{combo}b shows that the CO$_2$ photodesorption has a somewhat different thickness dependence at 60~K. The CO-from-CO$_2$ desorption dependence on thickness  is indistinguishable from a linear function and is fitted linearly. The total CO$_2$ desorption is fitted well, but not uniquely, by $2.2(\pm0.2)\times10^{-3} \times (1-e^{-(x/5.8(\pm1.2) \rm)})+2.2(\pm0.3)\times10^{-4}\times x$ molecules photon$^{-1}$, where $x$ is the ice thickness in monolayers. This formula indicates that photodesorption takes place deeper into the ice at higher temperatures and that the mean-free-path of CO becomes infinite.
 
\textit{Temperature}. At temperatures higher than 30~K the CO$_2$ photodesorption is initially fluence dependent, which is not the case for colder ices. This is further discussed below; here the constant yield reached after a fluence of $2.0\times10^{17}$ photons cm$^{-2}$ is used for comparison between the photodesorption yields at different temperatures. Figure \ref{combo}c shows the total CO$_2$ photodesorption yield dependence on temperature for different ice thicknesses. At 18 and 30~K both the total and the CO-from-CO$_2$ photodesorption yields are indistinguishable within the experimental uncertainties. Between 30 and 40~K the photodesorption yield jumps and above 40~K the photodesorption yield is again independent of temperature.

The build-up of CO molecules in the CO$_2$ ice is also temperature dependent. Above 30~K, the CO build-up is less than 50\% of the build-up at lower temperatures.

\textit{Photon flux.} Figure \ref{combo}d shows 4--6.5~ML CO$_2$ ices irradiated with different photon fluxes at 18 K. Between 1.1 and $8.3\times 10^{13}$ photons  cm$^{-2}$ s$^{-1}$ the CO$_2$ photodesorption yield in molecules photon$^{-1}$ is constant within the experimental uncertainties. The CO desorbing from the CO$_2$ ice is independent of the lamp flux as well. At 60~K the ice is irradiated at two different fluxes and, similarly to the colder ice, the photodesorption yield is constant (not shown). 

At 18~K, the produced CO ice increases throughout the experiment up to a fluence of $\sim10\times 10^{17}$ photons cm$^{-2}$. After $10\times 10^{17}$ photons cm$^{-2}$ the amount of CO ice reaches steady-state, which is only clearly visible in the experiment with the highest fluence. The observed independence of the CO-from-CO$_2$ photodesorption yield on CO ice content indicates that direct CO photodesorption from the formed CO ice plays a minor role during irradiation of the CO$_2$ ice.

\textit{Time and photon fluence.} Figure \ref{co2rate60k} shows the photodesorption of a 60~K, 7.4~ML CO$_2$ ice irradiated with $2.3\times10^{13}$ photons s$^{-1}$ cm$^{-2}$. At 60~K there is no measurable photodesorption during the first two hours or a fluence of $2\times10^{17}$ cm$^{-2}$. This can be compared with the 4.0 ML CO$_2$ ice at 18~K in Fig. \ref{des_test} (bottom), where photodesorption starts within a fluence of $10^{17}$ photons cm$^{-2}$. In the 60~K experiment, the total CO$_2$ photodesorption yield jumps to $3.0\times 10^{-3}$ molecules photon$^{-1}$ after a fluence of $2\times10^{17}$ photons cm$^{-2}$ and remains constant for the remainder of the experiment. This photodesorption delay is observed in all 5--7~ML ices at 40--60~K and also at all thicknesses between 3--11~ML at 60~K. The QMS measurements also show clear differences between the 18~K and the 60~K experiments (Fig. \ref{mscoco2}). At 60~K, the CO QMS signal is lower than at 18~K during the first 2$\times10^{17}$ photons cm$^{-2}$, corresponding to a yield $\sim1\times10^{-4}$ molecules photon$^{-1}$. After the first $10^{17}$ photons cm$^{-2}$ the CO signal increases rapidly with fluence.

To test whether this delay in photodesorption onset at 60~K is time or fluence dependent, a 5.8~ML ice is also irradiated at a 50\% lower flux. In this experiment the photodesorption according to the RAIRS begins after the same photon fluence, which occurs after twice the amount of time compared with the experiment at a higher flux.

\textit{Thermal annealing.} In one experiment, a 7.0~ML thick ice is deposited at 60~K and subsequently cooled down to 18~K before starting the irradiation. The CO-from-CO$_2$ photodesorption yield is $\sim1\times10^{-4}$ photon$^{-1}$, while only an upper limit of $5\times10^{-4}$ photon$^{-1}$ is derived for the total CO$_2$ photodesorption from the RAIRS measurements. This is significantly lower compared to both unannealed experiments at 18~K and to 60~K experiments (Fig. \ref{co2rate60k}). The CO-from-CO$_2$ QMS signal is similar to that from warm ices during the first $10^{17}$ photons cm$^{-2}$. The CO ice build-up, as measured from RAIRS, is similar to the un-annealed ice at 18~K.

\section{Discussion}

\subsection{CO and N$_2$ yields and mechanisms}

In \citet{Oberg07b} we concluded that CO photodesorbs from the top one or two ice layers at 15~K. Recent theoretical work shows that CO only desorbs from the absolute surface layer (Takahashi \& van Hemert, in prep.), which is supported by the new findings in this study. 

The experiments show that the photodesorption of CO depends on thermal annealing, such that annealing at a higher temperature results in a lower desorption yield.  The annealing most likely results in a more compact ice with a smaller effective surface area as well as stronger bound molecules. These two effects then add up to decrease the quantum efficiency of the photodesorption process. The reason for the linear dependence with temperature is unclear and may be coincidental. This behavior cannot be extrapolated to lower temperatures, since a 15~K ice should be amorphous. 

The new photodesorption experiments with N$_2$ further constrain the CO photodesorption mechanism. Consistent with \citet{Oberg07b} there is no evidence of direct N$_2$ photodesorption. The increase in photodesorption yield of N$_2$ when mixed with CO or grown in a monolayer on top indicates that $\sim$5\% of the UV photon absorptions of CO molecules result in the desorption of a neighboring molecule rather than the desorption of the originally excited molecule. The decreasing CO photodesorption in the mixed ice and the lack of initial CO desorption in the layered experiment also confirm that CO only desorbs from the top ice layer. 

\subsection{CO$_2$ yield and mechanism}

In contrast to CO and H$_2$O photodesorption \citep[Takahaski \& van Hemert, in prep.]{Andersson06, Andersson08}, the photodesorption mechanism of CO$_2$ has not been theoretically addressed yet. From the dependencies reported here, the mechanism may be constrained, however, and it is similar to that of H$_2$O. CO$_2$ photodissociates to CO+O, where the products have excess energy. This is followed by both reactions in the ice to form the observed CO$_3$, and recombination to CO$_2$. Some of these reaction products subsequently desorb.  

The flux independence of the CO$_2$ photodesorption has also previously been seen for H$_2$O and CO ices \citep[\"Oberg et al. accepted by ApJ]{Westley95a, Oberg07b}. This independence is expected for single photon processes, but not for multi photon processes or desorption induced by excess heat from the lamp. This is consistent with the suggested mechanism of dissociation fragments and recombined molecules traveling through the ices before desorbing.

The CO$_2$ photodesorption yield is clearly thickness dependent, which is in contrast to the CO photodesorption from surfaces only. This difference can be understood from the fact that before a desorption event occurs, the CO$_2$ molecule is dissociated into energetic products (whether concerned with the fragments or recombined molecules), which may travel through several monolayers of ice before they are quenched by the surrounding matrix. Assuming a homogeneous ice, the probability of a molecule with excess energy, from dissociation or recombination, reaching the ice surface and desorbing is only dependent on the excess energy, the diffusion properties of the molecule and the ice depth at which it receives its excess energy. The diffusion properties of a molecule are simplest described by the average distance the molecule travels through the ice before being stopped. Defining $l$ as the average distance traveled by a molecule before quenched by the surrounding ice, the fraction of particles with excess energy that will move through a slab of ice of thickness $x$ is expected to be proportional to $e^{-x/l}$, assuming uniform photochemistry throughout the ice and that the direction the molecule travels is independent of ice depth. Integrating over the ice depth from 0 to $x$, the total desorption of particles between 0 and $x$ is then proportional to $1-e^{-x/l}$.

We find that this type of expression describes the photodesorption at low temperatures well, with an average travel distance or mean-free-path of 2.9~ML for the CO$_2$ photodesorption and 4.6~ML for the CO fragment desorption. The two different values for the mean-free-paths have large uncertainties and it is not clear that the difference is significant. The values are however consistent with the different sizes of CO and CO$_2$, since the larger molecule is expected to be less mobile in the ice. The different mean-free-paths may also be partly due to different mechanisms, i.e. dissociation versus recombination, through which the CO fragment and the recombined CO$_2$ molecule receive excess energy.

The different mean-free-paths for CO and CO$_2$ desorption from CO$_2$ ice can also be consistent with a different desorption mechanism of CO$_2$ molecules -- momentum transfer to a surface CO$_2$ molecule from a smaller fragment originating in an underlying layer. This is observed in simulations by \citet{Andersson08} as an equally important photodesorption pathway for H$_2$O compared to desorption of the recombined molecule. The ice thickness dependence of this kind of process remains to be explored, since it requires more complex models than the simple mean-free-path model presented here. 

Other desorption pathways of the CO fragment can however be ruled out from the experiments presented here. The desorbed CO molecules do not originate in photodesorption of CO ice, since the yield does not increase with an increasing fraction of CO in the ice -- this fraction never reaches equilibrium during most low temperature experiments -- while the CO QMS signal does. In addition, the underlying substrate seems to have no influence on the desorption yield; hence substrate mediated processes are excluded as well.

At temperatures above the pure CO desorption temperature of $\sim$30~K, the increased photodesorption yield is most likely due to the increased mobility of CO$_2$ and CO in the ice. This is seen from the longer mean-free-path for desorbing CO$_2$ molecules and the infinite mean-free-path of CO. The latter points to a very high mobility of CO in the ice at these temperatures, which results in that most of the produced CO thermally desorbs once formed through CO$_2$ dissociation. It is also shown by the onset of O$_2$ photodesorption (Fig. \ref{co2_ms}). The desorption of O$_2$ is less than CO at any temperature (Section 3.2.1), but its mere presence shows that at 60~K oxygen atoms are very mobile and produced abundantly.

A curious and not yet fully understood feature is the delayed onset of both CO and CO$_2$ photodesorption from CO$_2$ ices at 40--60~K. Since it is a fluence rather than a time effect it is probably due to a re-structuring of the ice induced by UV irradiation. This restructuring is indicated by a change in the infrared spectral profile (not shown), which seems more pronounced for the warmer ices compared to the colder ones. This proposed re-structuring may also be responsible for the apparent initial increase in CO$_2$ in Fig. \ref{co2rate60k}. Ice band strengths depend on ice structure \citep[e.g.][]{Oberg07a} and this increase in CO$_2$ signal is most likely due to structural changes upon UV irradiation, modifying the strength of the CO$_2$ ice band, rather than more CO$_2$ molecules adsorbing. Similar changes in band strength have been previously observed in CO  and in H$_2$O ices bombarded by ions \citep{Loeffler05, Gomis04}

The importance of ice structure for the photodesorption yield is seen for CO as well, as discussed above. For CO$_2$ this importance is further shown by the experiment on a thermally annealed ice, where the photodesorption yield is $<40$\% of the yield of an amorphous ice. This is in contrast with the increased photodesorption yield for warm (40--60~K) ices, which are expected to have similar ice structures as the annealed ice. This new ice structure probably has a smaller effective ice surface compared to the 18~K amorphous ice, explaining the low photodesorption yield of the annealed ice. At temperatures above 40~K this decrease in surface does not affect the yield due to the high mobility of molecules in the ice. In summary, the CO$_2$ photodesorption yield is both temperature and structure dependent.

\subsection{Astrophysical significance}

\subsubsection{CO and N$_2$} 

The relevance of CO photodesorption at 15 K is discussed in detail in \citet{Oberg07b}. The results here show that the previously reported yield is dependent on temperature and annealing between 15 and 27 K. The astrophysical significance of this dependence is probably limited, however. Because the desorption temperature and annealing temperature are similar, only a minor fraction of the pure CO ice will ever be annealed in astrophysical environments. Therefore the photodesorption yield derived at 15 K may be used for most purposes, also where a real temperature gradient exists.

While a temperature gradient may not significantly reduce the CO photodesorption yield, a layer of N$_2$ ice on top of the CO ice does. This layer may form either through later freeze-out or through selective photodesorption as shown in the experiments. Because of the peculiar photodesorption mechanism of CO ice, this study shows that a single monolayer of N$_2$ ice initially reduces the CO photodesorption yield with more than 80\%. After a UV fluence of $8.5\times10^{17}$ photons cm$^{-2}$ (corresponding to $\sim$300 years of irradiation at cloud edges and $\sim$3 million years in cloud cores), the yield has increased to 25\% of the pure CO ice photodesorption yield, indicating that 25\% of the N$_2$ layer is desorbed. This probably happens through CO co-desorption, since the photodesorption yield of pure N$_2$ is too low to account for the observed desorption. N$_2$ co-desorption is also observed in a mixed CO:N$_2$ ice where N$_2$ photodesorbs at a yield of $3\times10^{-4}$ photon$^{-1}$. This is still a very low photodesorption yield compared to e.g. the CO photodesorption yield and its astrophysical significance is doubtful. Other non-thermal desorption mechanisms such as cosmic ray induced spot heating will likely be more important.

\subsubsection{CO$_2$}

At low temperatures and for thick ices the CO$_2$ photodesorption yield is $2.3(\pm1.4)\times10^{-3}$ photon$^{-1}$, which is almost identical to the CO photodesorption yield. Of the desorbed ice more than 50\% desorbs in the form of CO$_2$, while the remaining 20--50\% desorbs as CO. At higher temperatures some CO$_2$, up to 5\%, also desorbs as O$_2$. This means that similar CO$_2$ and CO abundances are expected in regions where photodesorption dominates, assuming that the photodesorption yields of the two molecules are not much different for astrophysical ice morphologies compared to the pure ices studied here. This remains to be investigated, especially for the astrophysically relevant cases of CO ice on top of CO$_2$ ice, and CO:CO$_2$ and CO$_2$:H$_2$O mixtures \citep{Pontoppidan08}. Based on our results we expect the recombined and thus energetic CO$_2$ molecule to penetrate at least as many CO ice layers during photodesorption as the 10 ML of CO$_2$ ice observed in this study, since CO forms a more loosely bound ice compared to CO$_2$. The behavior of CO$_2$ in a H$_2$O matrix is more difficult to predict. Hence, until such a laboratory study exists we recommend to use the equation presented here for CO$_2$ photodesorption, with the possible modification of taking into account that only a fraction of the ice is CO$_2$ ice.

In the regions in clouds and disks where ices begin to form, the ice thickness dependence of the CO$_2$ photodesorption needs to be taken into the account. As soon as the grain is covered with less than 10 ML of CO$_2$ ice, the CO$_2$ photodesorption yield decreases considerably with ice thickness according to $1.2\times10^{-3}\times(1-e^{-x/2.9})+1.1\times10^{-3}\times(1-e^{-x/4.6})$ at 18~K, where $x$ is the ice thickness and the two parts in the yield expression are due to CO$_2$ and CO desorption, respectively, during the irradiation of a CO$_2$ ice.

Heated or thermally annealed, i.e. heated and subsequently cooled down, CO$_2$ ice is observed toward many high- and low-mass protostars  \citep{Gerakines99, Pontoppidan08}. Because of the irreversibility of the infrared spectroscopic signature of heated CO$_2$ ice, the two cannot be easily distinguished. At temperatures higher than the pure CO sublimation line at 25--30~K, the CO$_2$ photodesorption yield increases to $2.2\times10^{-3}\times(1-e^{-x/5.8})+0.22\times10^{-3}\times x$ molecules photon$^{-1}$ for the CO$_2$ and the CO-fragment desorption, respectively. This results in a CO$_2$ yield increase of at most a factor of two. In addition, possible annealing may decrease the photodesorption yield to a value that is less than 40\% of the cold amorphous CO$_2$ ice yield. The uncertainty in the photodesorption yield is thus an additional factor of 2--3 in thermally processed regions.

\subsubsection{CO$_2$ astrophysical model}

CO$_2$ is not detected directly in the radio regime due to its lack of a permanent dipole moment. It is instead traced by HCO$_2^+$ toward e.g. the protostar L 1527 IRS. As a test case we use a simple model to investigate whether the recently observed HCO$_2^+$ toward L 1527 IRS may be explained by photodesorption of CO$_2$ ice. \citet{Sakai08} estimated the gas phase CO$_2$ abundance to be $>2.9\times10^{-7}$ with respect to H$_2$, by using a simple reaction scheme for the CO$_2$ to HCO$_2^+$ chemistry and assuming that the CO$_2$ is extended over the beam size of the IRAM 30 m telescope. \citet{Furlan08} report a CO$_2$ ice abundance of $5\times10^{-6}$ with respect to H$_2$ in L 1527 IRS, which is used in our model. This abundance is almost an order of magnitude lower than what is observed toward a large sample of low-mass protostars \citep{Pontoppidan08} and may be underestimated. The effect of increasing the CO$_2$ fractional abundance to $\sim3\times10^{-5}$ is discussed below.

To simplify the calculation we assume that the average temperature is below 30~K and that the hydrogen density is constant throughout the envelope. At this low temperature thermal desorption of CO$_2$ is negligible and therefore the equilibrium gas phase abundance of CO$_2$ is dependent only on the UV photodesorption and the freeze-out rates. We also make the approximation that the total CO$_2$ abundance is constant throughout the envelope, since observations show that CO$_2$ forms in ices at the edges of clouds and does not increase deeper into the cloud \citep{Pontoppidan06}. The UV field is composed of the interstellar radiation field of $10^8$ photons cm$^{-2}$ s$^{-1}$, which is attenuated with $A_{\rm V}$, and the UV photons produced inside the cloud by cosmic rays. For a typical galactic cosmic ray flux, the resulting UV photon flux is of the order of $10^4$ photons cm$^{-2}$ s$^{-1}$ with a factor of 3 uncertainty \citep{Shen04}. Using our derived photodesorption yields and the estimated freeze-out rate of CO$_2$ we calculate the steady-state gas abundance of CO$_2$ as a function of $A_{\rm V}$ for small and large grains. 

The photodesorption rates of CO$_2$ molecules from grain surfaces in molecules s$^{-1}$ due to external and cosmic ray induced UV photons, respectively, is described by

\begin{equation}
R_{ \rm PD-ISRF}=I_{\rm ISRF-FUV}e^{ \rm -\gamma A_V}Y_{ \rm pd}\left(\pi a_{ \rm gr}^2 \right), 
 \label{eqn1}
\end{equation}

\begin{equation}
R_{ \rm PD-CR}=I_{\rm CR-FUV}Y_{ \rm pd}\left(\pi a_{ \rm gr}^2 \right),
 \label{eqn2}
\end{equation}

\begin{equation}
Y_{ \rm pd}=0.0012(1-e^{-x/2.9}),
 \label{eqn3}
\end{equation}

\begin{equation}
x=n_{ \rm CO_2-ice}/\left(1\times10^{15}\times\pi a_{ \rm gr}^2 n_{ \rm gr}\right),
 \label{eqn4}
\end{equation}

\noindent and

\begin{equation}
n_{ \rm CO_2-ice}=5\times10^{-6}n_{\rm H} - n_{ \rm CO_2-gas},
 \label{eqn7}
\end{equation}

\noindent where $I_{\rm PD-ISRF}$ is the strength of the external irradiation field with energies 6-13.6 eV and $I_{\rm PD-FUV}$ is the strength of the UV field due to cosmic rays. $\gamma$ is a measure of UV extinction relative to visual extinction, which is $\sim$2 for small interstellar grains and $<$0.6 for grains of a few $\mu$m \citep{Roberge91,VanDishoeck06}, and $a_{ \rm gr}$ is the grain radius. $Y_{\rm pd}$ is the experimentally determined CO$_2$ photodesorption yield for temperatures below 30 K (Eq. \ref{eqn3}). $x$ is the ice thickness in monolayers, which is defined in terms of the CO$_2$ ice abundance ($n_{\rm CO_2-ice}$ in cm$^{-3}$), grain surface ($\pi a_{ \rm gr}^2 n_{ \rm gr}$ in cm$^2$ cm$^{-3}$) and amount of molecules per monolayer ($1\times10^{15}$ cm$^{-2}$) in Eq. \ref{eqn4}.  Equation \ref{eqn7} states the relationship between the gas and ice phase CO$_2$ abundances ($n_{\rm CO_2-gas}$ and $n_{ \rm CO_2-ice}$) when the total CO$_2$ abundance is $5\times10^{-6}n_{\rm H}$ and n$_{\rm H}$ is set to $1\times10^{4}$ cm$^{-3}$.

The accretion rate of CO$_2$ molecules $R_{\rm acc}$ is a product of the molecular velocity, the grain surface, the sticking coefficients and the gas phase abundance of CO$_2$ according to

\begin{equation}
R_{acc \rm}=-4.57\times 10^4 \left(\frac{T}{m_{CO_2 \rm}}\right)^\frac{1}{2}\left(\pi a_{gr \rm}^2 \right)S n_{CO_2-gas \rm}.
 \label{eqn6}
\end{equation}

\noindent In Eq. \ref{eqn6} the gas temperature $T$ is set to 10 K, $m_{\rm CO_2}$ is the CO$_2$ mass in atomic mass units,  and $S$ is the sticking coefficient, which is assumed to be 1 \citep{Bisschop06}.

\begin{figure}
\resizebox{\hsize}{!}{\includegraphics{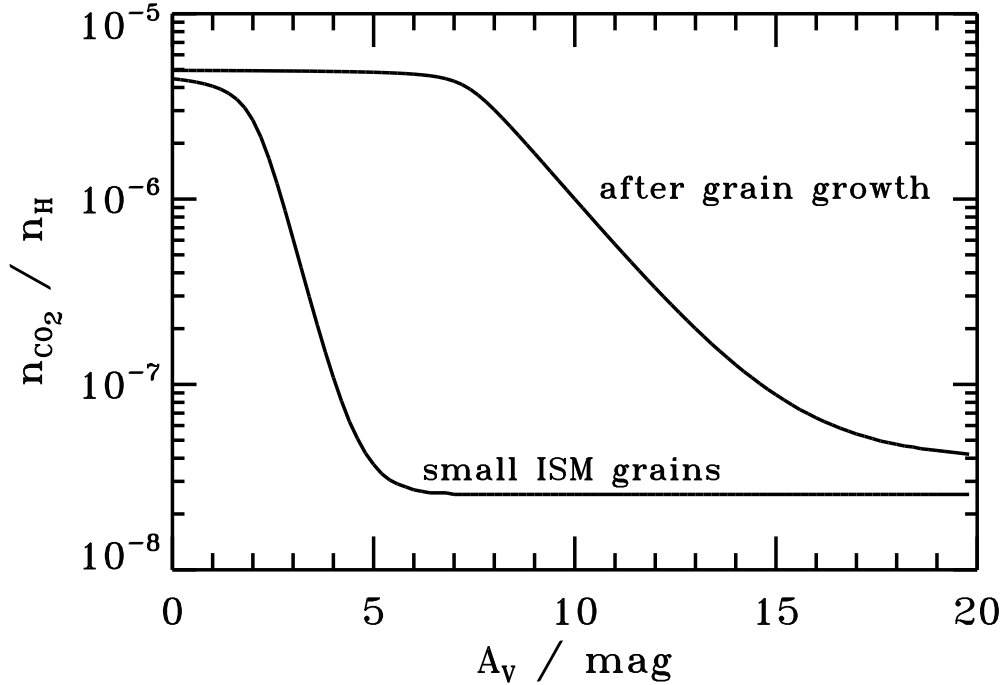}}
\caption{The CO$_2$ gas phase abundance with respect to the total hydrogen (H+H$_2$) column density assuming a total (gas + ice) CO$_2$ abundance of $5\times10^{-6}$. The two tracks assume classical 0.1~$\mu$m grains and grain growth up to a few $\mu$m, respectively.}
\label{co2astro}
\end{figure}

At steady-state the total photodesorption rate (i.e. the sum of Eqs. \ref{eqn1} and \ref{eqn2})  is equal to the accretion rate. Figure \ref{co2astro} shows the resulting steady-state CO$_2$ gas phase abundance as a function of $A_{\rm V}$ for small classical 0.1 $\mu$m grains and after grain growth to a few $\mu$m. At $A_{\rm V}$ $<$3 mag, CO$_2$ gas is photodissociated and the model is not valid there. Deep into the envelope, the gaseous CO$_2$ abundance due to cosmic ray induced photodesorption is $\sim2 \times10^{-8}$, which is about an order of magnitude lower than the observed CO$_2$ abundance. For small grains the external UV light also probably does not penetrate deep enough into the envelope to increase the average abundance significantly. On average about 1\% of the total CO$_2$ abundance is kept in the gas phase through photodesorption. If the CO$_2$ ice abundance is increased to $\sim3\times10^{-5}$, the CO$_2$ gas abundance increases by an order of magnitude at low extinctions. Deep into the cloud the abundance only increases by a factor of two, since the photodesorption rate only depends on the ice thickness up to $\sim$10 ~ML. In cloud and envelope material dominated by small grains, photodesorption probably does not explain gaseous CO$_2$ abundances of 5--10\%, as observed toward L 1527 IRS, unless there is a strong internal UV source and the UV is scattered into the cavities created by the outflow \citep{Spaans95} where it can photodesorb material in the cavity walls. Indeed, L1527 IRS is well known for its prominent outflow and X shaped cavity wall on scales comparable to the IRAM 30m beam \citep{MacLeod94,Hogerheijde98}.

In disks, or in general where grain growth has occurred, the picture changes dramatically. Millimeter observations of outer disks show grain growth up to mm size \citep{Rodmann06, Lommen07}, and the UV irradiation may then penetrate deep into the disk releasing a high fraction of the CO$_2$ ice into the gas phase. At an A$_{V \rm}$ of 10 mag the CO$_2$  fractional abundance is now $\sim 1\times10^{-6}$ or 20\% of the total CO$_2$ abundance. Similarly to the case with small grains, increasing the CO$_2$ ice abundance only increases the CO$_2$ gas abundance substantially at low extinctions.

This simple model thus shows that the experimentally determined CO$_2$ photodesorption yield is high enough to release large amounts of CO$_2$ ice into the gas phase if moderate grain growth has occurred. Even with small grains and no internal UV source except that produced by cosmic rays, photodesorption may keep up to $\sim$1\% of the total CO$_2$ abundance in the gas phase deep into cloud cores. This model stresses, once again, that photodesorption of ices needs to be taken into account when modeling gas-grain interactions.

\section{Conclusions}

   \begin{enumerate}
      \item The CO photodesorption yield is temperature dependent between 15 and 27 K, which is described empirically by $2.7\times10^{-3}-(T-15)\times1.7\times10^{-4}$ molecules photon$^{-1}$. The anti-correlation between yield and temperature is probably due to ice re-structuring into a more compact configuration -- the observed linearity may be coincidental, however. For most astrophysical applications the yield measured at 15 K is appropriate to use. 
\item The CO photodesorption is initially reduced by more than 80\% when the CO ice is covered by 1 ML of N$_2$ ice and decreases with UV fluence when mixed with N$_2$ due to surface build-up of N$_2$ ice, confirming that CO only desorbs from the ice surface.
\item N$_2$ co-desorbs with CO at 16 K in an ice mixture and in a layered ice with a yield of $3\times10^{-4}$ molecules photon$^{-1}$.
\item A CO$_2$ photodesorption event starts with the photodissociation of a CO$_2$ molecule into CO and O. The fragments either desorb directly or react and recombine to form CO$_2$ and CO$_3$ before desorbing. The CO$_3$ yield is however less than 1\% and the two main desorption products are CO and CO$_2$.
\item The CO$_2$ photodesorption yield is thickness dependent at all temperatures between 18 and 60 K. At 18--30 K the yield is well described by $1.2\times10^{-3}\times(1-e^{-x/2.9})+1.1\times10^{-3}\times(1-e^{-x/4.6})$, and at 40--60 K by $2.2\times10^{-3}\times(1-e^{-x/5.8})+0.22\times10^{-3}\times x$ molecules photon$^{-1}$, where $x$ is the ice thickness in monolayers. The first part in each yield equation is due to desorbing CO$_2$ molecules and the second part to desorbing CO molecules.
\item The thickness dependence of CO$_2$ photodesorption is understood from a mean-free-path perspective, where the different excited fragments travel a different average distance through the ice before being stopped. At higher temperatures, this mean free path increases due to increased mobility of molecules in the ice.
\item A simple model of an envelope using the observed CO$_2$ abundance in L 1527 IRS shows that CO$_2$ photodesorption can maintain CO$_2$ fractional abundances up to $1\times10^{-6}$ in the gas phase at A$_V\sim10$ mag after moderate grain growth and $2-3\times10^{-8}$ using small ISM grains. At lower extinctions the photodesorption is higher due to the external irradiation field and a high fraction of the total CO$_2$ ice abundance is maintained in the gas phase.
   \end{enumerate}

\begin{acknowledgements}
The authors wish to thank Stefan Andersson and Herma Cuppen for stimulating discussions. Funding is provided by NOVA, the Netherlands Research School for Astronomy, a grant from the European Early Stage Training Network ('EARA' MEST-CT-2004-504604) and a Netherlands Organization for Scientific Research (NWO) Spinoza grant.
\end{acknowledgements}

\bibliographystyle{aa}

\begin{thebibliography}{47}
\expandafter\ifx\csname natexlab\endcsname\relax\def\natexlab#1{#1}\fi

\bibitem[{{Andersson} {et~al.}(2006){Andersson}, {Al-Halabi}, {Kroes}, \& {van
  Dishoeck}}]{Andersson06}
{Andersson}, S., {Al-Halabi}, A., {Kroes}, G.-J., \& {van Dishoeck}, E.~F.
  2006, \jcp, 124, 064715

\bibitem[{{Attard} \& {Barnes}(2004)}]{adsorption}
{Attard}, G. \& {Barnes}, C. 2004, {Surfaces} ({Oxford Science Publications}),
  1--5

\bibitem[{{Bergin} {et~al.}(2002){Bergin}, {Alves}, {Huard}, \&
  {Lada}}]{Bergin02}
{Bergin}, E.~A., {Alves}, J., {Huard}, T., \& {Lada}, C.~J. 2002, ApJL, 570,
  L101

\bibitem[{{Bergin} {et~al.}(2001){Bergin}, {Ciardi}, {Lada}, {Alves}, \&
  {Lada}}]{Bergin01}
{Bergin}, E.~A., {Ciardi}, D.~R., {Lada}, C.~J., {Alves}, J., \& {Lada}, E.~A.
  2001, ApJ, 557, 209

\bibitem[{{Bisschop} {et~al.}(2006){Bisschop}, {Fraser}, {{\"O}berg}, {van
  Dishoeck}, \& {Schlemmer}}]{Bisschop06}
{Bisschop}, S.~E., {Fraser}, H.~J., {{\"O}berg}, K.~I., {van Dishoeck}, E.~F.,
  \& {Schlemmer}, S. 2006, A{\&}A, 449, 1297

\bibitem[{{Boogert} \& {Ehrenfreund}(2004)}]{Boogert04}
{Boogert}, A.~C.~A. \& {Ehrenfreund}, P. 2004, in ASP Conf. Ser. 309:
  Astrophysics of Dust, ed. A.~N. {Witt}, G.~C. {Clayton}, \& B.~T. {Draine},
  547--572

\bibitem[{Brewer \& Wang(1972)}]{Brewer72}
Brewer, L. \& Wang, J. L.-F. 1972, The Journal of Chemical Physics, 56, 759

\bibitem[{{Cottin} {et~al.}(2003){Cottin}, {Moore}, \&
  {B{\'e}nilan}}]{Cottin03}
{Cottin}, H., {Moore}, M.~H., \& {B{\'e}nilan}, Y. 2003, \apj, 590, 874

\bibitem[{{Dartois} {et~al.}(2003){Dartois}, {Dutrey}, \&
  {Guilloteau}}]{Dartois03}
{Dartois}, E., {Dutrey}, A., \& {Guilloteau}, S. 2003, A{\&}A, 399, 773

\bibitem[{{Dominik} {et~al.}(2005){Dominik}, {Ceccarelli}, {Hollenbach}, \&
  {Kaufman}}]{Dominik05}
{Dominik}, C., {Ceccarelli}, C., {Hollenbach}, D., \& {Kaufman}, M. 2005, ApJL,
  635, L85

\bibitem[{{Eidelsberg} {et~al.}(1992){Eidelsberg}, {Rostas}, {Breton}, \&
  {Thieblemont}}]{Eidelsberg92}
{Eidelsberg}, M., {Rostas}, F., {Breton}, J., \& {Thieblemont}, B. 1992, \jcp,
  96, 5585

\bibitem[{Fuchs {et~al.}(2006)Fuchs, Acharyya, Bisschop, \"Oberg, van
  Broekhuizen, Fraser, Schlemmer, van Dishoeck, \& Linnartz}]{Fuchs06}
Fuchs, G.~W., Acharyya, K., Bisschop, S.~E., {et~al.} 2006, Faraday
  Discussions, 133, 331

\bibitem[{{Furlan} {et~al.}(2008){Furlan}, {McClure}, {Calvet}, {Hartmann},
  {D'Alessio}, {Forrest}, {Watson}, {Uchida}, {Sargent}, {Green}, \&
  {Herter}}]{Furlan08}
{Furlan}, E., {McClure}, M., {Calvet}, N., {et~al.} 2008, \apjs, 176, 184

\bibitem[{{Gerakines} \& {Moore}(2001)}]{Gerakines01}
{Gerakines}, P.~A. \& {Moore}, M.~H. 2001, Icarus, 154, 372

\bibitem[{{Gerakines} {et~al.}(1996){Gerakines}, {Schutte}, \&
  {Ehrenfreund}}]{Gerakines96}
{Gerakines}, P.~A., {Schutte}, W.~A., \& {Ehrenfreund}, P. 1996, \aap, 312, 289

\bibitem[{{Hogerheijde} {et~al.}(1998){Hogerheijde}, {van Dishoeck}, {Blake},
  \& {van Langevelde}}]{Hogerheijde98}
{Hogerheijde}, M.~R., {van Dishoeck}, E.~F., {Blake}, G.~A., \& {van
  Langevelde}, H.~J. 1998, \apj, 502, 315

\bibitem[{{Hudgins} {et~al.}(1993){Hudgins}, {Sandford}, {Allamandola}, \&
  {Tielens}}]{Hudgins93}
{Hudgins}, D.~M., {Sandford}, S.~A., {Allamandola}, L.~J., \& {Tielens},
  A.~G.~G.~M. 1993, \apjs, 86, 713

\bibitem[{{L{\'e}ger} {et~al.}(1985){L{\'e}ger}, {Jura}, \& {Omont}}]{Leger85}
{L{\'e}ger}, A., {Jura}, M., \& {Omont}, A. 1985, A{\&}A, 144, 147

\bibitem[{{Lommen} {et~al.}(2007){Lommen}, {Wright}, {Maddison},
  {J{\o}rgensen}, {Bourke}, {van Dishoeck}, {Hughes}, {Wilner}, {Burton}, \&
  {van Langevelde}}]{Lommen07}
{Lommen}, D., {Wright}, C.~M., {Maddison}, S.~T., {et~al.} 2007, \aap, 462, 211

\bibitem[{{MacLeod} {et~al.}(1994){MacLeod}, {Avery}, \& {Harris}}]{MacLeod94}
{MacLeod}, J.~M., {Avery}, L.~W., \& {Harris}, A. 1994, \jrasc, 88, 265

\bibitem[{Mason {et~al.}(2006)Mason, Dawes, Holtom, Mukerji, Davis, Sivaraman,
  Kaiser, Hoffmann, \& Shaw}]{Mason06}
Mason, N.~J., Dawes, A., Holtom, P.~D., {et~al.} 2006, Faraday Discussions,
  133, 1

\bibitem[{Moll {et~al.}(1966)Moll, Clutter, \& Thompson}]{Moll65}
Moll, N.~G., Clutter, D.~R., \& Thompson, W.~E. 1966, The Journal of Chemical
  Physics, 45, 4469

\bibitem[{{Mu{\~n}oz Caro} \& {Schutte}(2003)}]{Munozcaro03}
{Mu{\~n}oz Caro}, G.~M. \& {Schutte}, W.~A. 2003, A{\&}A, 412, 121

\bibitem[{{{\"O}berg} {et~al.}(2007){{\"O}berg}, {Fuchs}, {Awad}, {Fraser},
  {Schlemmer}, {van Dishoeck}, \& {Linnartz}}]{Oberg07b}
{{\"O}berg}, K.~I., {Fuchs}, G.~W., {Awad}, Z., {et~al.} 2007, \apjl, 662, L23

\bibitem[{{{\"O}berg} {et~al.}(2005){{\"O}berg}, {van Broekhuizen}, {Fraser},
  {Bisschop}, {van Dishoeck}, \& {Schlemmer}}]{oberg05}
{{\"O}berg}, K.~I., {van Broekhuizen}, F., {Fraser}, H.~J., {et~al.} 2005,
  ApJL, 621, L33

\bibitem[{{Pi{\'e}tu} {et~al.}(2007){Pi{\'e}tu}, {Dutrey}, \&
  {Guilloteau}}]{Pietu07}
{Pi{\'e}tu}, V., {Dutrey}, A., \& {Guilloteau}, S. 2007, ArXiv Astrophysics
  e-prints

\bibitem[{{Pontoppidan}(2006)}]{Pontoppidan06}
{Pontoppidan}, K.~M. 2006, \aap, 453, L47

\bibitem[{{Pontoppidan} {et~al.}(2008){Pontoppidan}, {Boogert}, {Fraser}, {van
  Dishoeck}, {Blake}, {Lahuis}, {{\"O}berg}, {Evans}, \&
  {Salyk}}]{Pontoppidan08}
{Pontoppidan}, K.~M., {Boogert}, A.~C.~A., {Fraser}, H.~J., {et~al.} 2008,
  \apj, 678, 1005

\bibitem[{{Roberge} {et~al.}(1991){Roberge}, {Jones}, {Lepp}, \&
  {Dalgarno}}]{Roberge91}
{Roberge}, W.~G., {Jones}, D., {Lepp}, S., \& {Dalgarno}, A. 1991, \apjs, 77,
  287

\bibitem[{{Roberts} {et~al.}(2007){Roberts}, {Rawlings}, {Viti}, \&
  {Williams}}]{Roberts07}
{Roberts}, J.~F., {Rawlings}, J.~M.~C., {Viti}, S., \& {Williams}, D.~A. 2007,
  \mnras, 382, 733

\bibitem[{{Rodmann} {et~al.}(2006){Rodmann}, {Henning}, {Chandler}, {Mundy}, \&
  {Wilner}}]{Rodmann06}
{Rodmann}, J., {Henning}, T., {Chandler}, C.~J., {Mundy}, L.~G., \& {Wilner},
  D.~J. 2006, A{\&}A, 446, 211

\bibitem[{{Sakai} {et~al.}(2008){Sakai}, {Sakai}, {Aikawa}, \&
  {Yamamoto}}]{Sakai08}
{Sakai}, N., {Sakai}, T., {Aikawa}, Y., \& {Yamamoto}, S. 2008, \apjl, 675, L89

\bibitem[{{Shen} {et~al.}(2004){Shen}, {Greenberg}, {Schutte}, \& {van
  Dishoeck}}]{Shen04}
{Shen}, C.~J., {Greenberg}, J.~M., {Schutte}, W.~A., \& {van Dishoeck}, E.~F.
  2004, A{\&}A, 415, 203

\bibitem[{{Slanger} \& {Black}(1978)}]{Slanger78}
{Slanger}, T.~L. \& {Black}, T. 1978, The Journal of Chemical Physics, 68, 1844

\bibitem[{{Spaans} {et~al.}(1995){Spaans}, {Hogerheijde}, {Mundy}, \& {van
  Dishoeck}}]{Spaans95}
{Spaans}, M., {Hogerheijde}, M.~R., {Mundy}, L.~G., \& {van Dishoeck}, E.~F.
  1995, \apjl, 455, L167+

\bibitem[{{Sternberg} {et~al.}(1987){Sternberg}, {Dalgarno}, \&
  {Lepp}}]{Sternberg87}
{Sternberg}, A., {Dalgarno}, A., \& {Lepp}, S. 1987, ApJ, 320, 676

\bibitem[{{Thrower} {et~al.}(2008){Thrower}, {Burke}, {Collings}, {Dawes},
  {Holtom}, {Jamme}, {Kendall}, {Brown}, {Clark}, {Fraser}, {McCoustra},
  {Mason}, \& {Parker}}]{Thrower08}
{Thrower}, J.~D., {Burke}, D.~J., {Collings}, M.~P., {et~al.} 2008, \apj, 673,
  1233

\bibitem[{{Tielens} \& {Charnley}(1997)}]{Tielens97}
{Tielens}, A.~G.~G.~M. \& {Charnley}, S.~B. 1997, Origins of Life and Evolution
  of the Biosphere, 27, 23

\bibitem[{{Tielens} \& {Hagen}(1982)}]{Tielens82}
{Tielens}, A.~G.~G.~M. \& {Hagen}, W. 1982, \aap, 114, 245

\bibitem[{{Turner} {et~al.}(1999){Turner}, {Terzieva}, \& {Herbst}}]{Turner99}
{Turner}, B.~E., {Terzieva}, R., \& {Herbst}, E. 1999, \apj, 518, 699

\bibitem[{{van Dishoeck}(2006)}]{Vandishoeck06b}
{van Dishoeck}, E.~F. 2006, Proceedings of the National Academy of Science,
  103, 12249

\bibitem[{van Dishoeck {et~al.}(2006)van Dishoeck, Jonkheid, \& van
  Hemert}]{VanDishoeck06}
van Dishoeck, E.~F., Jonkheid, B., \& van Hemert, M.~C. 2006, Faraday
  Discussions, 133, 231

\bibitem[{{Watanabe} {et~al.}(2003){Watanabe}, {Shiraki}, \&
  {Kouchi}}]{Watanabe03}
{Watanabe}, N., {Shiraki}, T., \& {Kouchi}, A. 2003, \apjl, 588, L121

\bibitem[{{Westley} {et~al.}(1995{\natexlab{a}}){Westley}, {Baragiola},
  {Johnson}, \& {Baratta}}]{Westley95a}
{Westley}, M.~S., {Baragiola}, R.~A., {Johnson}, R.~E., \& {Baratta}, G.~A.
  1995{\natexlab{a}}, Nature, 373, 405

\bibitem[{{Westley} {et~al.}(1995{\natexlab{b}}){Westley}, {Baragiola},
  {Johnson}, \& {Baratta}}]{Westley95b}
{Westley}, M.~S., {Baragiola}, R.~A., {Johnson}, R.~E., \& {Baratta}, G.~A.
  1995{\natexlab{b}}, \planss, 43, 1311

\bibitem[{{Willacy} \& {Langer}(2000)}]{Willacy00}
{Willacy}, K. \& {Langer}, W.~D. 2000, ApJ, 544, 903

\bibitem[{{Willacy} \& {Millar}(1998)}]{Willacy98}
{Willacy}, K. \& {Millar}, T.~J. 1998, \mnras, 298, 562

\end{thebibliography}

\end{document}